\documentclass[onecolumn,nofootinbib,notitlepage]{revtex4-2}
\usepackage{subcaption}
\usepackage{float}   
\usepackage{natbib}
\usepackage[T1]{fontenc}
\usepackage[utf8]{inputenc}
\usepackage{amssymb,mathtools,graphicx}
\mathtoolsset{
	above-shortintertext-sep=-0.5\belowdisplayshortskip,
	below-shortintertext-sep=-0.5\abovedisplayshortskip,
	above-intertext-sep=0pt, 
	below-intertext-sep=0pt, 
}
\usepackage{bm,soul}
\usepackage{dsfont,bm,BOONDOX-cal}
\usepackage{hyperref}
\hypersetup{
	colorlinks   = true, 
	urlcolor     = blue, 
	linkcolor    = blue, 
	citecolor   = red 
}
\usepackage[scr=boondoxo,scrscaled=1.05]{mathalfa}
\usepackage{bookmark}
\bookmarksetup{
	numbered,
	open
}
\usepackage[resetlabels]{multibib}
\newcites{I}{References for the Introduction [I]}
\newcites{Th}{References for the Commentary [C]}
\usepackage{tikz}
\usetikzlibrary{calc}
\usepackage{pgf}

\newcommand{\ti}{t_{\mathfrak{0}}}
\newcommand{\tf}{t_{\mathfrak{1}}}
\colorlet{darkred}{red!85!black}
\colorlet{lightred}{red!25!white}
\colorlet{darkblue}{blue!90!black}
\colorlet{darkgreen}{green!50!black}
\begin{document}
\title{E. Schr\"odinger's 1931 paper ``On the Reversal of the Laws of Nature'' [``\"Uber die Umkehrung der Naturgesetze'', \textit{Sitzungsberichte der preussischen Akademie der Wissenschaften, physikalisch-mathematische Klasse}, \emph{8} N9 144-153]}
\author{Introduction and Commentary  by:}
\affiliation{}
 \author{Rapha\"el Chetrite}
\affiliation{CR CNRS, Laboratoire Dieudonn\'e,	Universit\' e de Nice Sophia-Antipolis, Nice, France}
\email{raphael.chetrite@unice.fr}
\author{Paolo Muratore-Ginanneschi}
\affiliation{Department of Mathematics and Statistics, University of Helsinki,	P.O. Box 68, 00014 Helsinki, Finland}
\email{paolo.muratore-ginanneschi@helsinki.fi}
\author{Kay Schwieger}
\affiliation{iteratec GmbH, Stuttgart, Baden-W\"urttemberg, Germany}
\email{kayschwieger@googlemail.com}
\author{Translated by: Paolo Muratore-Ginanneschi} \author{ Kay Schwieger}

\date{\today}

\begin{abstract}
We present an English translation of Erwin Schr\"odinger's paper on ``On the Reversal of the Laws of Nature‘'. In this paper Schr\"odinger analyses the idea of time reversal of a diffusion process. Schr\"odinger's paper acted as a prominent source of inspiration for the  works of Bernstein on reciprocal processes and of Kolmogorov on time reversal properties of Markov processes and detailed balance. The ideas outlined by Schr\"odinger also inspired the development of  probabilistic interpretations of quantum mechanics by F\'enyes, Nelson and others as well as the notion of  ``Euclidean Quantum Mechanics''   as probabilistic analogue of quantization. 
In the second part of the paper Schr\"odinger discusses the relation between time reversal and statistical laws of physics. We emphasize in our commentary the relevance of Schr\"odinger's intuitions for contemporary developments in statistical nano-physics. 
\end{abstract}

\maketitle

\section*{{\Large Introduction}}

Erwin Schr\"odinger had the rare privilege to be elected to  the Prussian Academy  of Science in February 1929, about one year and a half after his appointment to the chair of theoretical physics at the University of Berlin \citeI[I]{IMoo1989}. At the moment of his election, at the age of forty-two, Schr\"odinger was the youngest member of the Academy. Among physicists, other members of the Academy were Max Planck, who proposed Schr\"odinger's membership, Max von Laue, Walther Nernst and Albert Einstein who held a special Academy professorship.  
During their common years in Berlin, Einstein and Schr\"odinger became good personal friends \citeI[I]{IMoo1989}.

Schr\"odinger presented  ``On the Reversal of  the Laws of Nature'' to the Academy in March 1931. The intellectual context of the paper was marked by the intense debate on the interpretation of  quantum mechanics \citeI[I]{IWhZu1983}. As Schr\"odinger himself stated in \S~3 of the paper his ``\emph{current concern}'' was to point out the existence of  a classical probabilistic structure such that the probability density is given by ``\emph{the product of a certain solution of}'' [the forward diffusion equation] ``\emph{and a certain solution of}'' [the backward diffusion equation] and thus presenting ``\emph{a striking analogy with  quantum mechanics}'' (\S~4). 

The observation of this analogy initiated parallel, and somewhat intertwined, lines of research aiming at either finding classical probabilistic analogues of quantum mechanics or more directly a classical probabilistic interpretation of quantum mechanics. Already in 1933, Reinhold F\"urth showed  \citeI[I]{IFur1933} (see \citeI[I]{IPePMG2020} for a translation) what in current mathematical language could be phrased as the existence of uncertainty relations satisfied by the variance of a martingale of a diffusion process times the variance of the martingale's ``current velocity'' \citeI[I]{INel2001}. After the second world war, F\"urth's work became the starting point of the ``stochastic mechanics'' program proposed by Imre F\'enyes \citeI[I]{IFen1952} and Edward Nelson \citeI[I]{INel1985, INel2001} as a probabilistic interpretation of quantum mechanics. A similar program was also pursued by Masao Nagasawa \citeI[I]{INag1964,INag1993} and Robert Aebi \citeI[I]{IAeb1996}. The collection \citeI[I]{IFaGrSiBrCa2006} offers a recent appraisal of the state of the art focusing on Nelson's contributions. 

In a spirit perhaps closer to Schr\"odinger's original idea,  ``Euclidean Quantum Mechanics''  \citeI[I]{IZam1997} applies modern developments of stochastic calculus of variations and optimal control theory (see \citeI[I]{IZam2009,ILeRoZa2014} for recent surveys) to study how ``\emph{relations between quantum physics and classical probability theory}'' may ``\emph{lead to new theorems in regular quantum mechanics}'' \citeI[I]{IZam1997}.

From the mathematical angle, and in particular the theory of Markov processes, the importance of the ideas put forward by Schr\"odinger was immediately realized. 
Sergei Bernstein discussed the contents of Schr\"odinger's paper in his address to the International Conference of Mathematicians held in Z\"urich in 1932 \citeI[I]{IBer1932}. 
Andre\v{i} Kolmogorov starts his 1936 ``On the Theory of Markov Chains'' paper \citeI[I]{IKol1936} discussing the interest of studying time reversal of Markov processes for the ``\emph{analysis of the reversibility of the  statistical laws of nature}'' explicitly referring to Schr\"odinger's paper. Kolmogorov's 1937 paper on detailed balance \citeI[I]{IKol1937} has the title ``On the Reversibility of the Statistical Laws of Nature'' thus clearly resonating the title of Schr\"odinger's paper.

Our interest in presenting an English translation, so far missing to the best of our knowledge, of  ``On the Reversal of  the Laws of Nature'' is more directly motivated by the second part of the paper, especially \S~6, where Schr\"odinger discusses fluctuations and time reversal in classical statistical physics. 

The advent of nano-manipulations has made possible  accurate laboratory observations of systems, natural and artificial, operating in contact with highly fluctuating environments. Fundamental questions concerning non-equilibrium statistical laws can be now posed in well controlled experimental setups \citeI[I]{IPePi2020}.
Theoretical analysis then requires a careful overhaul of the meaning of time reversal and dissipation in systems whose evolution laws can only be defined in a statistical sense see e.g. \citeI[I]{IMaReMo2000,IChGa2008}.

 In particular, we want to draw the attention of readers to the analogy between the probabilistic mass transport problem discovered by Schr\"odinger and the optimal control problem associated to the derivation of Landauer's bound in non-equilibrium statistical nano-physics \citeI[I]{ILeOrPoSn2019}. In the remarkable paper \citeI[I]{ILan1961} Rolf Landauer introduced his conjecture that only logically irreversible information processes are fundamentally linked to irreversible, i.e. finite dissipation, thermodynamic processes. Landauer's conjecture has been the object of many refinements and criticisms see \citeI[I]{ILeRe2003,ILeOrPoSn2019} for overviews. The existence of a finite lower bound to the average cost of erasing a bit of memory has  been ascertained in several recent experiments \citeI[I]{IBeArPeCiDiLu2012,IKoSaSaYoKuSoMoAlPe2013,IDaPeBaCiBe2021}.

To delve a bit deeper into the analogy, we devote our commentary after the translation to first explaining, drawing also from \citeI[I]{IAeb1996,IAeb1996b}, how the problem considered by Schr\"odinger can be rephrased as an optimal control problem. Next, we  try to give a glimpse into the modern, infinite dimensional, counter-part of Schr\"odinger's optimal control problem \citeI[I]{IWak1989,IWak1991,IBla1992,IDaPr1991,IPaWa1991,IPav2003,IMik2004,IPaTi2010,IChGePa2016,IThTo2012}. Finally, we turn to the  analogy with the mathematical problem of proving the existence in the average sense of  Landauer's  bound to the energy dissipated in a non-equilibrium transformation between target states.
We refer to the lecture notes \citeI[I]{IGaw2013} for a masterly introduction to this subject.

We strove to write the commentary in a self-contained way. Our aim is to give readers a concise  overview of Schr\"odinger's paper from the standpoint of current developments in non-equilibrium statistical mechanics.  

We conclude this introduction with a very much needed apology to all authors whose work we were not able to give proper visibility in our necessarily limited set of references.

\renewcommand{\refname}{References for the Introduction}
\bibliographyI{reversibility_intro} 
\bibliographystyleI{abbrv}

\begin{center}
	***
\end{center}

\section*{{\Large On the Reversal of the Laws of Nature}}

\subsection*{Introduction}
\label{sec:intro}

If the probability to be in the interval $(x;x+dx)$ at time $t_0$ 
\begin{align}
	w(x,t_{\mathfrak{0}}) \; \mathrm{d} x
	\nonumber
\end{align}  
for a particle diffusing or performing a Brownian motion is given,
\begin{align}
	w(x,t_{\mathfrak{0}}) = w_{\mathfrak{0}}(x),
	\nonumber
\end{align}
then it is precisely the  solution $ w(x,t) $ for $ t\,>\,t_{\mathfrak{0}} $ of the diffusion equation
\begin{equation}
	\label{1}
	D \frac{\partial^2 w}{\partial x^2} = \frac{\partial w}{\partial t}
\end{equation}
that becomes equal at $t=t_{\mathfrak{0}}$ to the given function $w_{\mathfrak{0}}(x)$.
There is an extensive literature on problems of this kind, including
many possible variations and complications suggested by special experimental
arrangements and observation methods whereby the system in question
does not need to be a diffusing particle at all but, for example, the electromechanical
meter needle in the experimental setup devised by  K.~W.~F.~Kohlrausch to measure
Schweidler oscillations, and equation (\ref{1}) is replaced by its generalization,
the so-called Fokker[-Planck] partial differential equation for the relevant system subject
to some random influences \cite{Fokker1914,Planck1917}.

Such systems also give rise to a class of problems in probability theory which has been hitherto neglected or has received little attention, 
and which is already of interest  from the purely
mathematical side since the answer is not specified by a
single solution of a Fokker[-Planck] equation but rather, as we will show, by the product of the solutions of two adjoint equations, and with time boundary conditions
imposed not on an individual solution but on the product. 

From the physics side there is a close relation to the class of problems that
M.~von~Smoluchowski \cite{Smo1913,Smo1914a,Smo1914b,Smo1915a,Smo1915b,Smo1915c,Smo1916} has uncovered in his latest beautiful works
on the waiting and return times of very unlikely configurations in systems of
diffusing particles. The conclusions, which we draw in \S~6, can already be
read off from the results of Smoluchowski, but occasion once more a sense of surprise in their sharp paradoxes. Furthermore (\S~4) there are remarkable
analogies with quantum mechanics that seem to me worth considering.

\subsection*{\S~1}

The simple example that I want to deal with here is the following.
Let the probability to find the particle in a certain position
be assigned not only at time $t_{o}$ but also at a second time instant $t_{\mathfrak{1}}> t_{\mathfrak{0}}$:
\begin{align}
	w(x,t_{\mathfrak{0}})=w_{\mathfrak{0}}(x);
	\hspace{1.0cm}
	w(x,t_{\mathfrak{1}})=w_{\mathfrak{1}}(x).
	\nonumber
\end{align}
What is the probability for \emph{intermediate times}, i.e., for any $t$ such that
\begin{align}
	t_{\mathfrak{0}}\leq t \leq t_{\mathfrak{1}}. 
	\nonumber
\end{align}
Obviously $w(x,t)$ is \emph{not} solution of (\ref{1}) since any solution of (\ref{1}) is
already fully specified  at any later time by its \emph{initial value}. Nor is $w$  solution of the \emph{adjoint} equation 
\begin{align}
	\label{2}
	D \, \frac{\partial^2 w}{\partial x^2} = - \frac{\partial w}{\partial t},
\end{align}
as this solution, in turn, would be fully specified at any \emph{prior} time by its final value $w_1(x)$. Is the question somehow ill posed? This is certainly not the case.
One recognizes this by considering a special case which we want to present in first place.
Let us suppose that we have detected the particle at time $t_{\mathfrak{0}}$ in $x_{\mathfrak{0}}$ and at
time $t_{\mathfrak{1}}$ in $x_{\mathfrak{1}}$ ($w_{\mathfrak{0}}$ and $w_{\mathfrak{1}}$ are then ``Spitzenfunktionen''
\footnote{
	literally: spike functions. In modern language Dirac $\delta$ functions.
} 
respectively sharply peaked at $x=x_{\mathfrak{0}}$ and $x=x_{\mathfrak{1}}$). An auxiliary observer has observed the position of the particle 
at time $t$ without, however, reporting us the result. The question is then: which 
probabilistic inferences can we draw from our two observations for the intervening
observations of our assistant? 

The answer is simple. I introduce the notation $g(x,t)$ for the well-known \emph{fundamental solution} of 
(\ref{1}):
\begin{align}
	\label{3}
	g(x,t)=\frac{1}{\sqrt{ 4\,\pi\,D\,t}} \, e^{-\frac{x^2}{4\,D\,t}}.
\end{align}  
This is the probability density at position $x$ and time $t>0$ if  the particle 
starts from $x=0$ at time $t=0$. Now I let the particle start many times, say $N$-times, from $x=x_{\mathfrak{0}}$.
Of such $N$ experiments, I single out the ones 
for which the particle is in $(x_{\mathfrak{1}},x_{\mathfrak{1}}+\mathrm{d}x)$ at time $t_{\mathfrak{1}}$. Their number is
\begin{align}
	n_{\mathfrak{1}} = N\,g(x_{\mathfrak{1}}-x_{\mathfrak{0}}, t_{\mathfrak{1}}-t_{\mathfrak{0}}) \, \mathrm{d} x_{\mathfrak{1}}.
	\nonumber
\end{align}
Of these I single out again the ones 
for which
\emph{1.}  the particle is in $(x, x+\mathrm{d}x)$ at time $t$ and then 
\emph{2.}  the particle is in $(x_{\mathfrak{1}},x_{\mathfrak{1}}+\mathrm{d}x_{\mathfrak{1}})$ at time $t_{\mathfrak{1}}$
The number of these experiments is
\begin{align}
	n = N \, g(x-x_{\mathfrak{0}}, t-t_{\mathfrak{0}}) \mathrm{d} x \; g(x_{\mathfrak{1}}-x,t_{\mathfrak{1}}-t) \; \mathrm{d} x_{\mathfrak{1}} .
	\nonumber
\end{align}
The probability we are after is clearly the ratio $n/n_{\mathfrak{1}}$, i.e.,
\begin{equation}
	\label{4}
	w(x,t) = \frac{g(x-x_{\mathfrak{0}}, t-t_{\mathfrak{0}}) \; g(x_{\mathfrak{1}}-x,t_{\mathfrak{1}}-t)}{g(x_{\mathfrak{1}}-x_{\mathfrak{0}}, t_{\mathfrak{1}}-t_{\mathfrak{0}})}.
\end{equation} 
This is the solution for the special case when at time $t_{\mathfrak{0}}$ and time $t_{\mathfrak{1}}$
the position of the particle is known \emph{with certainty}.

\subsection*{\S~2}

We now consider the general case. 
The experimental setup is as follows.
We let  a large number $N$ of particles start at time $t_{\mathfrak{0}}$, namely
\begin{align}
	\label{5}
	N\,w_{\mathfrak{0}}(x_{\mathfrak{0}})\,\mathrm{d}x_{\mathfrak{0}}
\end{align}   
from the interval $(x_{\mathfrak{0}},x_{\mathfrak{0}}+\mathrm{d}x_{\mathfrak{0}})$. We observe that at time
$t_{\mathfrak{1}}$
\begin{equation}
	\label{6}
	N\,w_{\mathfrak{1}}(x_{\mathfrak{1}})\,\mathrm{d}x_{\mathfrak{1}}
\end{equation}
arrived in the interval $(x_{\mathfrak{1}},x_{\mathfrak{1}}+\mathrm{d}x_{\mathfrak{1}})$. (Incidental remark:
this observation may be more or less \emph{surprising}, and as such renders the outcome of
our series of experiments more or less \emph{exceptional}. The reason is 
that instead of (\ref{6}) one would \emph{expect}:
\begin{equation}
\label{6'}	
	\tag{\ref{6}'}
	N\,\mathrm{d}x_{\mathfrak{1}} \int_{-\infty}^{\infty} w_{\mathfrak{0}}(x_{\mathfrak{0}})\,g(x_{\mathfrak{1}}-x_{\mathfrak{0}},t_{\mathfrak{1}}-t_{\mathfrak{0}}) \; \mathrm{d} x_{\mathfrak{0}} .
\end{equation}
This is not, however, our concern here. We assume that the distributions 
(\ref{5}) and (\ref{6}) are \emph{actually realized} and we have to draw 
conclusions based on \emph{this  fact}.)

The solution of this more general problem is considerably more difficult than
in the special case previously considered. If we knew how many of the particles 
(\ref{5}) contribute to (\ref{6}), then we would have to multiply 
\emph{this number} by (\ref{4}) and then to integrate $x_{\mathfrak{0}}$ and $x_{\mathfrak{1}}$ from $-\infty$
to $+\infty$. Determining the aforementioned number is the main task.

We divide the $x$-axis in cells of equal size which, for simplicity's sake, we take of unit length. We call $a_{k}$ the number (\ref{5}) which at time $t_{\mathfrak{0}}$ starts
from the $k$-th cell, $b_{l}$ the number (\ref{6}) which at time $t_{\mathfrak{1}}$ lands in the
$l$-th cell. Let $g_{k l}$ be the a priori probability for a particle starting from the
$k$-th cell to arrive to the $l$-th cell, i.e., $g_{k \,l}$ is an appropriate notation for $g(x_{\mathfrak{1}}-x_{\mathfrak{0}},t_{\mathfrak{1}}-t_{\mathfrak{0}})$ in the 
present case and satisfies $g_{l k}=g_{k l}$. Finally, let $c_{k l}$ be the number of particles which  arrive \emph{into} the $l$-th \emph{from} the $k$-th cell. The following equations therefore apply
\begin{equation}
	\label{7}
	\left.
	\begin{array}{ll}
		\sum_{l}c_{k l}= a_{k} 
		&
		\quad \text{for any}\,k,
		\\[0.3cm]
		\sum_{k}c_{k l}= b_{l} 
		&
		\quad \text{for any}\,l.
	\end{array}
	\right\}
\end{equation}
Between the equations (\ref{7}) there is one and only one identity
which stems from
\begin{align}
	\label{8}
	\sum_{k}a_{k}=\sum_{l}b_{l}=N.
\end{align}
The matrix $c_{kl}$ is clearly \emph{not} given. The actually observed particle migration can come into being according to any of the $c_{k\,
	l}$-matrices compatible with (\ref{7}). In the limit $N=\infty$ (which
is of course always meant) it will be, however, correct to assume that the
actual migration will be realized with complete certainty by that $c_{k\, l}$-matrix
which attributes the largest probability to the migration. Even for fixed
$c_{k\,l}$, the actually observed particle migration can be realized in very
many different ways. One way is that one knew in which cell each individual
particle landed. This possible realization yields for the observed outcome the probability
\begin{align}
	\label{9}
	\prod_{k}\prod_{l}g_{k l}^{c_{k l}}.
\end{align} 
As mentioned above, there are, however, very many such equally probable possible realizations, specifically
\begin{align}
	\label{10}
	\prod_{k}\frac{a_{k}!}{\prod_{l}c_{k l}!}.
\end{align}
The product of (\ref{9}) and (\ref{10}) results in the total probability yielded for 
the observed outcome by a fixed choice of $c_{k l}$
\begin{align}
	\label{11}
	\prod_{k}a_{k}!\prod_{k}\prod_{l}\frac{g_{k l}^{c_{k l}}}{c_{k l}!}.
\end{align}
Now as usual, we look for that  $c_{k l}$ which maximizes (\ref{11})
under the constraints (\ref{7}). One easily finds
\begin{equation}
	\label{12}
	c_{k l} = g_{k l} \psi_{k} \phi_{l}.
\end{equation}
The  $\psi_{k}$'s and $\phi_{l}$'s are Lagrangian multipliers. They are determined
by the constraints
\begin{align}
	\label{13}
	\left.
	\begin{array}{ll}
		\psi_k \sum_l g_{k l} \phi_l = a_{k} \hspace{1.0cm}&\hspace{1.0cm}\mbox{for any}\,k,
		\\[0.3cm]
		\phi_l \sum_k g_{k l} \psi_k =b_{l}\hspace{1.0cm}&\hspace{1.0cm}\mbox{for any}\,l.
	\end{array}
	\right\}
\end{align}
Now we have to translate (\ref{12}) and (\ref{13}) back to the language of the continuum. $a_{k}$ and $b_{l}$ are specified by (\ref{5}) and (\ref{6}). $\psi_{k}$ and $\phi_{l}$ are functions of $x$, namely we shall set
\begin{align*}
	\psi_{k} &= \sqrt{ N}\,\psi(x_{\mathfrak{0}}) \mathrm{d}x_{\mathfrak{0}}
	&
	\phi_{l} &= \sqrt{ N}\,\phi(x_{\mathfrak{1}})\mathrm{d}x_{\mathfrak{1}}.
\end{align*}
Furthermore, $g_{k l}=g(x_{\mathfrak{1}}-x_{\mathfrak{0}},t_{\mathfrak{1}}-t_{\mathfrak{0}})$ holds. Hence
\begin{equation}
	\label{13'}
	\tag{\ref{13}'}
	\left.
	\begin{array}{l}
		\psi(x_{\mathfrak{0}})\int_{-\infty}^{\infty}g(x_{\mathfrak{1}}-x_{\mathfrak{0}},t_{\mathfrak{1}}-t_{\mathfrak{0}})\,\phi(x_{\mathfrak{1}})\mathrm{d}x_{\mathfrak{1}}=w_{\mathfrak{0}}(x_{\mathfrak{0}})
		\\[0.3cm]
		\phi(x_{\mathfrak{1}})\int_{-\infty}^{\infty}g(x_{\mathfrak{1}}-x_{\mathfrak{0}},t_{\mathfrak{1}}-t_{\mathfrak{0}})\,\psi(x_{\mathfrak{0}})\mathrm{d}x_{\mathfrak{0}}=w_{\mathfrak{1}}(x_{\mathfrak{1}})
	\end{array}
	\right\}
\end{equation}
and
\begin{equation}
	\label{12'}
	\tag{\ref{12}'}
	c(x_{\mathfrak{0}}, x_{\mathfrak{1}}) \mathrm{d} x_{\mathfrak{0}} \, \mathrm{d} x_{\mathfrak{1}}
	=
	N\,g(x_{\mathfrak{1}}-x_{\mathfrak{0}},t_{\mathfrak{1}}-t_{\mathfrak{0}}) \, \psi(x_{\mathfrak{0}}) \,\phi(x_{\mathfrak{1}}) \, \mathrm{d} x_{\mathfrak{0}} \, \mathrm{d} x_{\mathfrak{1}}
\end{equation}
is the desired number of particles which diffuse from
$(x_{\mathfrak{0}},x_{\mathfrak{0}}+\mathrm{d}x_{\mathfrak{0}})$ to $(x_{\mathfrak{1}},x_{\mathfrak{1}}+\mathrm{d}x_{\mathfrak{1}})$. If we
multiply (\ref{12'}) by (\ref{4}) and integrate over $x_{\mathfrak{0}}$ and $x_{\mathfrak{1}}$,
we then obtain (after dividing by $N$) the probability density at $x$ and time
$t$:
\begin{align}
	\label{14}
	w(x,t)=\int_{-\infty}^{\infty}g(x-x_{\mathfrak{0}},t-t_{\mathfrak{0}})\,\psi(x_{\mathfrak{0}})\mathrm{d}x_{\mathfrak{0}} \cdot \int_{-\infty}^{\infty}
	g(x_{\mathfrak{1}}-x,t_{\mathfrak{1}}-t)\,\phi(x_{\mathfrak{1}})\mathrm{d}x_{\mathfrak{1}}.
\end{align} 
\emph{This is the solution of the problem} expressed in terms of the solution of the integral 
system (\ref{13'}).

\subsection*{\S~3}
\label{sec:3}

The discussion of this pair of equations would be certainly interesting but
probably not simple because it is non-linear. The existence and uniqueness of
the solution (except perhaps for very tricky choices of $w_{\mathfrak{0}}$ and $w_{\mathfrak{1}}$) I
take for granted because of the reasonable question which in an unambiguous and
sharp manner leads to these equations. Our current concern is less how to
\emph{actually} construct $\psi$ and $\phi$ from given  $w_{\mathfrak{0}}$
and $w_{\mathfrak{1}}$ than the general form of $w(x,t)$. The latter is in fact extremely
transparent: \emph{the product of an arbitrary solution of \eqref{1} and an arbitrary solution of \eqref{2}}. Namely the first factor in \eqref{14}
is nothing else than an arbitrary solution of \eqref{1} distinguished by
$\psi(x_{\mathfrak{0}})$, its value distribution at time $t_{\mathfrak{0}}$. The same applies to the second factor
in \eqref{14} with respect to equation \eqref{2}. Furthermore, it is a
simple consequence of \eqref{1} and \eqref{2} that the product of
two solutions has a time independent $\int_{-\infty}^{\infty}\mathrm{d}x\dots$
\emph{preserving} normalization to unity if it was normalized to unity
at some time. (This restriction must be obviously imposed: one must
only use $ 2 $ solutions whose product has a \emph{finite value} of
$\int_{-\infty}^{\infty} \mathrm{d} x \dots$, so that it can be normalized to $ 1 $). And then within the time interval in which the product of the solutions remains regular one may choose arbitrarily \emph{any two} times $t_\mathfrak{0}$ and $t_\mathfrak{1}$ as the ones for which the probability density has been observed (of course observed to be precisely as given by the values in the product). Then the product yields the probability density for \emph{intermediate times}.

\subsection*{\S~4}
\label{sec:4}

The most interesting thing about result today is the striking analogy with
quantum mechanics. The existence of a certain relationship between the
fundamental equation of wave mechanics and the Fokker[-Planck] equation, as between the statistical concepts arising from both of them, have probably impressed anyone familiar enough with both circles of ideas. And yet, a closer inspection reveals
two very deep discrepancies. The first is that in the \emph{classical} theory
of random systems the probability density \emph{itself} obeys a linear
differential equation, whereas in wave mechanics this is the case for the so-called probability
amplitudes, from which all probabilities are formed \emph{bilinearly}. The
second discrepancy resides in the following fact: whilst in both cases the
differential equation is of \emph{first} order in time, the presence of a
factor $\sqrt{-1}$ confers to the wave equation a hyperbolic or, physically
stated, reversible character at variance with the parabolic-irreversible
character of the Fokker[-Planck] equation.

In both these points, the example considered above shows a much closer analogy with wave mechanics although it concerns a classical, originally irreversible system. As in wave mechanics, the probability density is given not by the solution of a single Fokker[-Planck] equation but by the product of two equations
differing only in the sign of the time variable. Thus the solution does not privilege any time direction either. If one exchanges $w_{\mathfrak{0}}(x)$ with $w_{\mathfrak{1}}(x)$, one
obtains precisely the reverse evolution of $w(x,t)$ between $t_{\mathfrak{0}}$ and $t_{\mathfrak{1}}$.
(In a certain sense, however, this fact also holds true for the simpler problem with just
a \emph{single} time boundary condition: if only the probability density at time $t_\mathfrak{0}$ is given  and nothing more, then the solution takes the same
value at time $t_{\mathfrak{0}}+t$ and $t_{\mathfrak{0}}-t$.)

Whether this analogy will prove useful to clarify notions in quantum mechanics I cannot foresee, yet. The aforementioned $\sqrt{-1}$ obviously constitutes, despite everything,  a very far reaching difference. I cannot restrain myself from quoting here some words of A.~S.~Eddington on the interpretation of quantum mechanics---obscure as they may be---which can be found on page 216f of his Gifford lectures \cite{Edd1928}
\begin{quote}
	The whole interpretation is very obscure, but it seems to depend on whether you are considering the probability \emph{after you know what has happened} or the probability for the purposes of prediction. The $\psi\psi^*$ is obtained by introducing two symmetrical systems of $\psi$ waves traveling in opposite directions in time; one of these must presumably correspond to probable inference from what is known (or is stated) to have been the condition at a later time.
\end{quote}

\subsection*{\S~5}
\label{sec:5}

We wish now to write (\ref{14}) in the form
\begin{align}
	\label{15}
	w(x,t)=\Psi(x,t)\,\Phi(x,t),
\end{align}
where we assume that $\Psi$ is a solution of \eqref{1}, $\Phi$ is a solution of \eqref{2} and the product $\Psi\,\Phi$ is normalized to 
unity:
\begin{align}
	\label{16}
	D\,\frac{\partial^2\Psi}{\partial{x}^{2}} 
	&= \frac{\partial \Psi}{\partial{t}} ,
	&
	D\,\frac{\partial^2\Phi}{\partial{x}^{2}} 
	&= -\frac{\partial \Phi}{\partial{t}} ,
	&
	\int_{-\infty}^{\infty} \Psi\,\Phi\, \mathrm{d} x
	&= 1.
\end{align}
Upon multiplying the first equation by $x\,\Phi$, the second by $-x\,\Psi$, and adding them, one gets
\begin{align}
	\nonumber
	\frac{\partial}{\partial t}\left(x\,\Phi\,\Psi\right)
	= D\,x\,\frac{\partial}{\partial x} \left(\Phi \frac{\partial \Psi}{\partial x} - \Psi \frac{\partial \Phi}{\partial x} \right).
\end{align}
Now compute $\int_{-\infty}^{\infty}\ldots$ and integrate by parts:
\begin{align}
	\nonumber
	\frac{\mathrm{d}}{\mathrm{d} t}\int_{-\infty}^{\infty} x \, w \, \mathrm{d} x 
&	= -D \, \int_{-\infty}^{\infty} \Bigl( \Phi \frac{\partial \Psi}{\partial x} - \Psi \frac{\partial \Phi}{\partial x} \Bigr) \, \mathrm{d} x.
	\nonumber\\
&= 2\,D\,\int_{-\infty}^{\infty} \Psi \, \frac{\partial \Phi}{\partial x} \, \mathrm{d} x .
\nonumber
\end{align}
On the left hand side there is the \emph{velocity} with which the \emph{barycenter} of the probability density moves. The integral on the right hand side, however, is constant  because:
\begin{gather*}
	\frac{\mathrm{d}}{\mathrm{d} t} \int_{-\infty}^\infty \Psi \, \frac{\partial \Phi}{\partial x} \, \mathrm{d} x
	= \int_{-\infty}^\infty \Bigl( \frac{\partial \Psi}{\partial t} \frac{\partial \Phi}{\partial x} + \Psi \frac{\partial^2 \Phi}{\partial x \partial t} \Bigr) \, \mathrm{d} x
	=
	\\
	= \int_{-\infty}^\infty \Bigl( \frac{\partial^2 \Psi}{\partial x^2} \frac{\partial \Phi}{\partial x} - \Psi \frac{\partial^3 \Phi}{\partial x^3} \Bigr) \, \mathrm{d} x
	= \int_{-\infty}^\infty \frac{\partial}{\partial x} \Bigl( \frac{\partial \Psi}{\partial x} \, \frac{\partial \Phi}{\partial x} - \Psi \frac{\partial^2 \Phi}{\partial x^2} \Bigr) \, \mathrm{d} x
	= 0.
\end{gather*}
\emph{The center of mass thus moves with constant velocity from its initial to
	its final position.}

In the special case when the initial and the final position of the particle are
known  sharply, equation \eqref{4}, one can furthermore state that the
\emph{maximum} of the probability moves uniformly from the initial to the
final position. This is because \eqref{4} is at any time a Gaussian
distribution, hence at any time the maximum and the mean value coincide.

\subsection*{\S~6}
\label{sec:6}

In a special case it is possible to specify immediately the solution of the pair of
integral equations (\ref{13'}). Namely, when the density $w_{\mathfrak{1}}$ prescribed at the
end of the time interval is precisely the one  to which the initial distribution $w_{\mathfrak{0}}$ evolves 
according to the free action of the diffusion equation (\ref{1}), i.e., if
\begin{align}
	\nonumber
	w_{\mathfrak{1}}(x_{\mathfrak{1}}) = \int_{-\infty}^{\infty} g(x_{\mathfrak{1}}-x_{\mathfrak{0}},t_{\mathfrak{1}}-t_{\mathfrak{0}})\,w_{\mathfrak{0}}(x_{\mathfrak{0}}) \, \mathrm{d} x_{\mathfrak{0}}.
\end{align}
Then obviously, one has to set
\begin{align*}
	\phi &\equiv 1;
	&
	\psi &\equiv w_{\mathfrak{0}}.
\end{align*}
$w(x,t)$ satisfies then \eqref{1} in the entire time interval. If one imagines it as the diffusion process of many particles then this is a
thermodynamically completely \emph{normal} diffusion process. 

But also,
conversely, when the \emph{initial distribution} $w_{\mathfrak{0}}$ is precisely the one
into which the final distribution $w_{\mathfrak{1}}$ would evolve during the time $t_{\mathfrak{1}}-t_{\mathfrak{0}}$
according to the free action of the normal (!) diffusion equation
\eqref{1}; or in other words: when the final distribution is prescribed in
such a way that it arises from the initial distribution following the
\emph{reversed} diffusion equation \eqref{2} in the time $t_{\mathfrak{1}}-t_{\mathfrak{0}}$; \emph{also in this case} the solution of \eqref{13'} is equally simple. In fact the assumption  then reads
\begin{align}
	\nonumber
	w_{\mathfrak{0}}(x_{\mathfrak{0}}) = \int_{-\infty}^{\infty} g(x_{\mathfrak{1}}-x_{\mathfrak{0}},t_{\mathfrak{1}}-t_{\mathfrak{0}}) \, w_{\mathfrak{1}}(x_{\mathfrak{1}}) \, \mathrm{d} x_{\mathfrak{1}},
\end{align} 
and the solutions of \eqref{13'} are
\begin{align*}
	\phi &\equiv w_{\mathfrak{1}};
	&
	\psi &\equiv 1.
\end{align*}
$w(x,t)$ then satisfies in the whole time interval the ''reversed'' equation \eqref{2}, 
the corresponding diffusion process is thermodynamically as abnormal as possible.
This, of course, occurs because of the odd boundary conditions,  but renders possible a
very interesting application to reality, namely to the \emph{way} extremely unlikely
exceptional states, that are  occasionally, even if extremely rarely, to be expected,  
occur in a system in thermodynamic equilibrium. 

Indeed, let us assume that  we have observed  the usual uniform distribution in a system of diffusing particles
 at time $t_{\mathfrak{0}}$ and a  substantial deviation therefrom at a later time $t_{\mathfrak{1}}$,  yet not so
substantial not to noticeably return to the uniform distribution after following the
diffusion law for a time $t_{\mathfrak{1}}-t_{\mathfrak{0}}$. In addition, we assume to know with
certainty that for intermediate times the system is left to itself in
unperturbed thermodynamic equilibrium or, in other words, that the observed
abnormal distribution is truly a spontaneous thermodynamic fluctuation phenomenon. If we were asked our opinion about the \emph{previous history}
that the observed strongly abnormal distribution could have probably had, then
we would have to reply that its first signs probably date back as long as it will take for its last traces to disappear; that from these
first signs an unfathomable swelling of the anomaly would have been
occasioned by  diffusion currents that  almost always almost exactly flowed
in the direction of the concentration gradient (\emph{upward} and not downward
slope) but, beside this sign difference, corresponded to the material
[diffusion] constant $D$: in brief that the anomaly was probably caused by a
precise time reversal of a normal diffusion process. Admittedly, this statement
about the likely previous history would be only a probabilistic judgment,
nevertheless it should,  in my opinion, be granted the same degree of ``almost
certainty'' as the corresponding statement about the likely future evolution,
i.e., about the normal diffusion process to be expected for $t > t_{\mathfrak{1}}$.

This statement, of course, must not lead to the misconception that a 
diffusion current in the \emph{direction} of the gradient and with magnitude
precisely corresponding to the diffusion constant $D$ would be \emph{in itself} much less
unlikely than any other biased current of arbitrary magnitude. Our probabilistic
conclusion is based not only on the diffusion mechanism but in an essential
manner on the knowledge of the strongly anomalous final state, which we assume
to have been actually observed. It turns out that it can always be attained in
an infinitely simpler manner and with an exceedingly larger probability by a
precise time reversal of the diffusion equation than by any other less radical
means. 

All the above may be without effort applied to arbitrary thermodynamic fluctuation phenomena
as soon as they substantially exceed the range  of normal fluctuations. The so-called irreversible laws of nature, if one interprets them statistically, do
\emph{actually} not privilege any time direction. This is because what they say in the particular case, depends only upon the time boundary
conditions at two ``cross sections'' ($t_{\mathfrak{0}}$ and $t_{\mathfrak{1}}$) and is completely symmetric with respect to
these cross sections without any special consequence
associated to their time ordering. This fact is only somewhat concealed inasmuch
we  in general consider only \emph{one} of the two ``cross sections'' as really
observed whilst for the other the reliable rule holds that if it is removed sufficiently far in time then one may assume  that the state of maximum
disorder or of maximum entropy applies there. That this rule is correct,
is actually very peculiar and, in my opinion, not logically deducible. But in
any case also this rule does not privilege any arrow of time inasmuch it  applies equally
in either of the two time directions  the second cross section is removed provided it 
is at sufficient time separation from the first.   

Incidentally, all of this was quite certainly already the explicit opinion of Boltzmann. In no other way can one understand for instance when he states the following at the end of his paper ``\"{U}ber die sogenannte H-Kurve'' \cite{Boltz1898} the following\footnote{
	See also \S~90 in \cite{Boltz1896}, furthermore \cite{Boltz1895,Boltz1897}; moreover compare the aforementioned works of Smoluchowski;
	among new authors G.N. Lewis in particular upholds the principle of
	``Symmetry of Time'' (e.\,g. \cite{Lewis1930} and elsewhere).
}:
\begin{quote}
	There is no doubt that we might as well conceive a world in which all natural processes can occur in  reversed order. 
	And yet a human being living in such a world would not have a different perception from us. 
	He would just refer to as future what we refer to as past.
\end{quote}
To those who regard as trivial and needless the extensive substantiation of this old thesis by means of
the diffusion processes which were so exhaustively studied in this context already by
Smoluchowski, I apologize. I will gladly subscribe to their opinion. But during discussions about these
matters I occasionally encountered considerable objections which made me unsure. It was suggested that
the laws governing  the \emph{emergence} through fluctuations of a strongly anomalous
state from a normal one are not nearly as strict as those governing its
\emph{disappearance};  rather that  a certain anomalous state, if one accumulates enough record of its rare occurrences through appropriately
long observation times, is \emph{relatively} frequently attained through a
\emph{completely} disordered process, which does \emph{not} correspond to the time reversal image of a normal process.

\subsection*{\S~7}

The considerations of the first three paragraphs can be applied with minor changes also to much more complicated cases: several spatial coordinates, variable diffusion 
coefficients, external forces which are arbitrary functions of  position. 
One always gets  a probability density in the form of the product of  solutions 
of two adjoint equations which in general not only differ in the sign of the time variable but also in 
other terms. One finds that
fundamental solutions (see equation (\ref{3}) above) of the adjoint equations enjoy the simple
(and certainly not new) property that they are obtained from one another by exchanging 
the coordinates of of the boundary conditions [des Aufpunkts und des Singularitaetspunkts] and by inverting the sign of time.
But I do not wish to analyze these points more closely before time tells if they can really lead to a better understanding of quantum mechanics.

\renewcommand{\refname}{Schr\"odinger's References}
\bibliography{reversibility}
\bibliographystyle{abbrv}
	
\begin{center}
	***
\end{center}

\section*{{\Large Commentary}}

The scope of these notes is first of all to explain the relation of the particle migration model of  section \S~2 of Schr\"odinger's paper 
with the modern theory of large deviations \citeTh[C]{CoTh2006,deHo2000,Ell2005,DeZe2009,Tou2009}.   Ahead of his time, Schr\"odinger  solves the particle migration model in the large deviation limit.  In doing so he identifies a quantifier of the divergence of the probability of a migration when the sample space is restricted to migrations between pre-assigned initial and final particle distributions from the probability of a particle migration when only the initial particle distribution is assigned whereas the final distribution is arbitrary. 
The quantifier turns out to be a relative entropy,  the Kullback--Leibler divergence \citeTh[C]{KuLe1951}, between a process connecting the two assigned probability densities and a reference process. 
Schr\"odinger uses the Kullback--Leibler divergence to formulate an optimal mass transport problem in the continuum limit \citeTh[C]{Vil2009}. 
Namely, equations (\ref{13'}) of Schr\"odinger's paper specify the minimizer of the Kullback--Leibler divergence between a reference Markov process and a second process whose transition probability evolves an initial assigned probability density into a target one, equally pre-assigned. 
Schr\"odinger explicitly constructs a probability density continuously interpolating between the assigned boundary conditions at the end of a time interval. 
The interpolating density admits a product decomposition reminiscent of Born's law in Quantum Mechanics.  
Schr\"odinger derives this result without explicitly introducing microscopic dynamics in the continuum limit.
Mainly drawing from \citeTh[C]{DaPr1991}, we show that the interpolating density can itself be directly regarded as the solution of a stochastic optimal control problem stemming from a microscopic formulation in terms of stochastic differential equations.

Next, we turn our attention to the notion of time reversal for Markov process introduced by Kolmogorov in  \citeTh[C]{Kol1936}. Kolmogorov considered his results ``\emph{in spite of their simplicity, to be new and not without interest for certain physical applications, in particular for the analysis of the reversibility of the statistical laws of nature, which Mr. Schr\"odinger  has carried out in the case of a special example}'' \citeTh[C]{Kol1936}. 

Finally, we briefly discuss the analogy between Schr\"odinger's  optimal control problem and the mathematical derivation of Landauer's bound (in the mean value sense) for diffusion processes. 

Overall, the scope of this commentary is to offer the reader a first brief overview, admittedly incomplete in spite of our effort, of the significance of Schr\"odinger's paper for current research from a non-equilibrium statistical physics perspective.

\section{From particle migration model to optimal control}
\label{sec:ld}

\setcounter{section}{1}
\setcounter{equation}{0}
\renewcommand{\theequation}{C.\arabic{equation}}

Our description of the particle migration model draws from \citeTh[C]{Aeb1996,Aeb1996b} which also provide further mathematical details and references.

\subsection{Formulation of Schr\"odinger's particle migration model}

\begin{minipage}{0.5\textwidth}

We suppose that 
\begin{enumerate}
	\item[-] two sets $ \left\{ A_{i} \right\}_{i=\mathfrak{1}}^{\mathfrak{n}} $ and $ \left\{ B_{i} \right\}_{i=\mathfrak{1}}^{\mathfrak{n}} $ of $\mathfrak{n}$ boxes 
	\item[-] $N$ particles initially randomly located in the set $ \left\{ A_{i} \right\}_{i=\mathfrak{1}}^{\mathfrak{n}} $ 
\end{enumerate}
are given. We want to move \emph{all} the particles from a given distribution in the first set of boxes to 
an assigned distribution in the second set. We suppose that each particle \emph{migration}
is an independent event which occurs with probability  
\begin{align}
	\nonumber
	g_{i\,j}=\mathrm{Pr}(\mbox{one particle migrates from}\,\,  A_{i}\,\, \mbox{to}\,\, B_{j}\,).
\end{align}
\end{minipage}
\begin{minipage}{0.50\textwidth}
			\begin{center}
	\begin{minipage}{0.80\textwidth}
		\begin{figure}[H]
			\nonumber
  \begin{tikzpicture}[%
	auto,
	block/.style={
		rectangle,
		draw=blue,
		thick,
		fill=blue!20,
		text width=5em,
		align=center,
		rounded corners,
		minimum height=2em
	},
	block1/.style={
		rectangle,
		draw=blue,
		thick,
		fill=blue!20,
		text width=5em,
		align=center,
		rounded corners,
		minimum height=2em
	},
	line/.style={
		draw,thick,
		-latex',
		shorten >=2pt
	},
	cloud/.style={
		draw=red,
		thick,
		ellipse,
		fill=red!20,
		minimum height=1em
	}
	]
		\path (-2,2) node[block] (A1) {}
	(-2,1) node[block] (A2) {}
	(-2,0) node[block] (A3) {}
	(-2,-1) node[block] (A4) { $ \vdots$}
	(-2,-2) node[block] (A5) { }
	(2,2) node[block] (B1) {}
	(2,1) node[block] (B2) {}
	(2,0) node[block] (B3) {}
	(2,-1) node[block] (B4) {$ \vdots$}
	(2,-2) node[block] (B5) {};
	\node  at (-2.7,2.1) {$ A_{\mathfrak{1}} $};
	\node  at (-2.7,1.1) {$ A_{\mathfrak{2}} $};
	\node  at (-2.7,0.17) {$ A_{\mathfrak{3}} $};
	\node  at (-2.7,-1.9) {$ A_{\mathfrak{n}} $};
	\node  at (2.7,2.1) {$ B_{\mathfrak{1}} $};
	\node  at (2.7,1.1) {$ B_{\mathfrak{2}} $};
	\node  at (2.7,0.17) {$ B_{\mathfrak{3}} $};
	\node  at (2.7,-1.9) {$ B_{\mathfrak{n}} $};
	\draw[-stealth,red] (A1.east) -- (B2.west);
	\draw [-stealth,red](A1.east) -- (B4.west);
	\draw[-stealth,red] (A2.east) -- (B1.west);
	\draw [-stealth,red](A2.east) -- (B2.west);
    \draw[-stealth,red] (A2.east) -- (B5.west);
\draw[-stealth,red] (A3.east) -- (B3.west);
\draw [-stealth,red](A4.east) -- (B1.west);
\draw[-stealth,red] (A4.east) -- (B5.west);
\draw[-stealth,red] (A5.east) -- (B4.west);
\draw[red,fill] (-2.4,2.2) circle[radius=3.0pt]; 
\draw[red,fill] (-2.0,2.2) circle[radius=3.0pt]; 
\draw[red,fill] (-1.6,2.2) circle[radius=3.0pt]; 
\draw[red,fill] (-1.2,2.2) circle[radius=3.0pt]; 
\draw[red,fill] (-2.4,1.8) circle[radius=3.0pt]; 
%
\draw[red,fill] (-2.4,1.2) circle[radius=3.0pt]; 
\draw[red,fill] (-2.0,1.2) circle[radius=3.0pt]; 
\draw[red,fill] (-1.6,1.2) circle[radius=3.0pt]; 
\draw[red,fill] (-1.2,1.2) circle[radius=3.0pt]; 
\draw[red,fill] (-2.4,0.8) circle[radius=3.0pt]; 
\draw[red,fill] (-2.0,0.8) circle[radius=3.0pt]; 
%
\draw[red,fill] (-2.4,0.2) circle[radius=3.0pt]; 
\draw[red,fill] (-2.0,0.2) circle[radius=3.0pt]; 
\draw[red,fill] (-1.6,0.2) circle[radius=3.0pt]; 
\draw[red,fill] (-1.2,0.2) circle[radius=3.0pt]; 
\draw[red,fill] (-2.4,-0.2) circle[radius=3.0pt]; 
\draw[red,fill] (-2.0,-0.2) circle[radius=3.0pt]; 
%
\draw[red,fill] (-2.4,-1.8) circle[radius=3.0pt]; 
\draw[red,fill] (-2.0,-1.8) circle[radius=3.0pt]; 
\draw[red,fill] (-1.6,-1.8) circle[radius=3.0pt]; 
\draw[red,fill] (-1.2,-1.8) circle[radius=3.0pt]; 
\draw[red,fill] (-2.4,-2.2) circle[radius=3.0pt]; 
%
\draw[red,fill] (2.2,2.0) circle[radius=3.0pt]; 
\draw[red,fill] (1.8,2.0) circle[radius=3.0pt]; 
\draw[red,fill] (1.4,2.0) circle[radius=3.0pt]; 
\draw[red,fill] (2.2,1.2) circle[radius=3.0pt]; 
\draw[red,fill] (1.8,1.2) circle[radius=3.0pt]; 
\draw[red,fill] (1.4,1.2) circle[radius=3.0pt]; 
\draw[red,fill] (1.4,0.8) circle[radius=3.0pt]; 
\draw[red,fill] (1.4,0.0) circle[radius=3.0pt]; 
\draw[red,fill] (1.8,0.0) circle[radius=3.0pt]; 
\draw[red,fill] (2.2,-1.8) circle[radius=3.0pt]; 
\draw[red,fill] (1.8,-1.8) circle[radius=3.0pt]; 
\draw[red,fill] (1.4,-1.8) circle[radius=3.0pt]; 
\draw[red,fill] (1.4,-1.8) circle[radius=3.0pt]; 
\draw[red,fill] (1.8,-2.2) circle[radius=3.0pt]; 
\end{tikzpicture}
\caption{Graphical illustration of the particle migration model}
\end{figure}
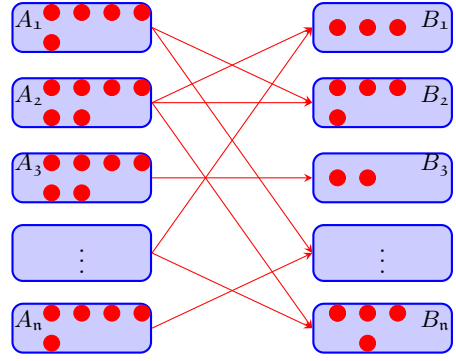
\end{minipage}
\end{center}
\end{minipage}
\vspace{0.2cm}

Particles can move in many ways. In other words, there can be multiple realizations of the particle migration process. We can put each realization 
in correspondence with the realization  of a random variable $ \mathscr{C}=\left\{ \mathscr{c}_{i \,j} \right\}_{i,j=1}^{\mathfrak{n}}$.    
The ensemble $\Gamma$ of the admissible values of $ \mathscr{C} $ consists of $\mathfrak{n}\,\times\,\mathfrak{n}$ \emph{square matrices $C =\{c_{i\,j}\}_{i,j=1}^{\mathfrak{n}}$ with integer elements} satisfying  the constraint
\begin{align}
	\label{me:normalization}
	\sum_{i,j=1}^{\mathfrak{n}} c_{i\,j}=N.
\end{align}
The interpretation of the matrix elements is
\begin{align}
	c_{i\,j}=\mbox{number of particles migrating from}\,A_{i}\,\mbox{to}\,B_{j}.
	\nonumber
\end{align}
The probability to sample the matrix $C$ from the ensemble $ \Gamma $ is then specified by the multinomial distribution i.e.
\begin{align}
	\nonumber
	\mathrm{Pr}(\mathscr{C}=C)=N!\prod_{i,j=\mathfrak{1}}^{\mathfrak{n}}\frac{g_{i\,j}^{c_{i\,j}}}{c_{i\,j}!}.
\end{align}
Furthermore, we may suppose that in each box $ A_{i} $ there are exactly $ a_{i} $  particles:
\begin{align}
		\label{me:constraint}
	\sum_{j=1}^{\mathfrak{n}}c_{i\,j}=a_{i}\,,\hspace{1.0cm}i=1,\dots,\mathfrak{n}
\end{align}
with the obvious constraint $\sum_{i=1}^{\mathfrak{n}}a_{i}=N$.
It is expedient to store the information about the marginal particle distribution in the boxes $ \left\{ A_{i} \right\}_{i=1}^{\mathfrak{n}} $ in a 
vector-valued random variable $ \widetilde{\mathscr{C}} $ whose components equal the sum over columns of $ \mathscr{C} $:
\begin{align}
	\widetilde{\mathscr{C}}_{i}=\sum_{j=1}^{\mathfrak{n}}\mathscr{c}_{i\,j}.
	\nonumber
\end{align}
The events fixing the marginal 
\begin{align}
	\left\{ \widetilde{\mathscr{C}}=\bm{a}\right\}=\bigcap_{i=1}^{\mathfrak{n}}\left\{  \sum_{j=1}^{\mathfrak{n}}\mathscr{c}_{i\,j}=a_{i}\right\}.
	\nonumber
\end{align}
also obey a multinomial distribution
\begin{align}
	\mathrm{Pr}(\widetilde{\mathscr{C}}=\bm{a})=\frac{N!}{\prod_{i=\mathfrak{1}}^{\mathfrak{n}}\,(a_{i}!)}\prod_{i=1}^{\mathfrak{n}}\left(\sum_{j=1}^{\mathfrak{n}} g_{i\,j}\right)^{a_{i}}.
	\label{me:multi}
\end{align}
We, therefore, recover Schr\"odinger's equation~(\ref{11}) in the form
\begin{align}
	\label{me:conditional}
	\mathrm{Pr}(\mathscr{C}=C\hspace{0.2cm}\,\big{|}\,\widetilde{\mathscr{C}}=\bm{a})=
	\left (\prod_{i=\mathfrak{1}}^{\mathfrak{n}}\,a_{i}!\right )\prod_{i,j=\mathfrak{1}}^{\mathfrak{n}}\frac{1}{c_{i\,j}!}\left(\frac{g_{i\,j}}{\sum_{j=1}^{\mathfrak{n}}g_{i\,j}}\right)^{c_{i\,j}}.
\end{align}
The matrix elements $ c_{i\,j} $'s on the right-hand side are now subject to the constraints (\ref{me:constraint}), whereas 
\begin{align}
	\mathrm{Pr}(\mbox{one particle's  arrival to}\,\,  B_{j}\,\, \mbox{under the \emph{condition} that it started from}\,\, A_{i})=\frac{g_{i\,j}}{\sum_{j=1}^{\mathfrak{n}}g_{i\,j}}\,\equiv\,g(j|i).
	\nonumber
\end{align}

\subsection{Large deviation and relative entropy }
\label{sec:me}

For large $ N $ we can estimate multinomial probabilities by means of Stirling's formula. To this effect we introduce the initial probability distribution $ \left\{ w_{\mathfrak{0}}(i) \right\}_{i=1}^{\mathfrak{n}} $ and relate it to the initial particle distribution via
\begin{align}
	a_{i}=N\,w_{\mathfrak{0}}(i).
	\nonumber
\end{align}
Similarly, we associate to each realization of $ \mathscr{C} $ an \emph{empirical probability distribution} $ \left\{ k_{i\,j} \right\}_{i,j=1}^{\mathfrak{n}} $ by setting
\begin{align}
	\nonumber
	c_{i \,j}= N\,k_{i\, j}
\end{align}
and, correspondingly, an empirical transition probability
\begin{align}
	k(j|i)=\frac{k_{i\, j}}{w_{\mathfrak{0}}(i)}.
	\nonumber
\end{align}
Upon retaining only leading order contributions in the large $ N $-limit, after some straightforward algebra we get  the \emph{large deviation}  asymptotics of the probability (\ref{me:multi})
\begin{align}
\label{me:ld}
	\mathrm{Pr}(\mathscr{C}=C\,\big{|}\,\bm{\widetilde{\mathscr{C}}}=\bm{a})\asymp \exp \left(-\,N\,\sum_{i \,j=1}^{\mathfrak{n}}k(j|i)\,w_{\mathfrak{0}}(i)\ln\frac{k(j|i)}{g(j|i)}\right).
\end{align}
The symbol $ \asymp $ emphasizes that in (\ref{me:ld}) we are neglecting sub-exponential corrections. This is the gist of any large deviation estimate. Indeed, an arbitrary random quantity $ \alpha_{N} $ depending upon a positive definite parameter $ N $ is said to satisfy a large deviation principle if  its probability distribution is amenable to the form
\begin{align}
	\mathrm{P}(\alpha_{N}=a) \asymp e^{-N\,I(a)}
	\nonumber
\end{align}
for $ N $ tending to infinity. 
The positive-definite function $ I $ is usually referred to as ``rate function'' or ``Cram\'er function''. A large deviation estimate implies an exponential decay of the probability distribution except when $ \alpha_{N} $ attains its typical value $ a_{\star} $ such that
\begin{align}
	I(a_{\star})=0.
	\nonumber
\end{align}
In the particular case of (\ref{me:ld}) the rate function coincides with the relative entropy or \emph{Kullback--Leibler  divergence} \mbox{\citeTh[C]{KuLe1951}} (see also \citeTh[C]{CoTh2006} ) between the empirical ($ \mathsf{K}=\left\{  k(j|i) \right\}_{i,j=1}^{\mathfrak{n}} $)  and the  a priori ($\mathsf{G}=\left\{ g(j|i) \right\}_{i,j=1}^{\mathfrak{n}} $) transition probabilities \emph{averaged} with respect to the initial particle distribution
\begin{align}
	\label{me:re}
	\mathrm{D}_{\scriptscriptstyle{KL}}(\mathsf{K}\|\mathsf{G})=\sum_{i \,j=1}^{\mathfrak{n}}k(j|i)\,w_{\mathfrak{0}}(i)\ln\frac{k(j|i)}{g(j|i)}.
\end{align}
The Kullback--Leibler  is a positive definite quantity measuring  how one probability distribution is different from a second, reference probability distribution. 
Namely, upon applying the elementary inequality 
\begin{align}
	\label{me:ineq}
	\ln(1/x) \ge 1-x \,,\hspace{1.0cm}\forall\,x\,>\,0
\end{align}
we  immediately obtain
\begin{align}
	\nonumber
	\mathrm{D}_{\scriptscriptstyle{KL}}(\mathsf{K}\|\mathsf{G})
	\geq 
	\sum_{i \,j=1}^{\mathfrak{k}} k(j|i) \, w_{\mathfrak{0}}(i) \left( 1 - \frac{g(j|i)}{k(j|i)} \right)
	= 0.
\end{align}
From these considerations it immediately follows that for (\ref{me:ld}) the typical value of the rate function corresponds to the case when the two conditional probabilities coincide.

Previous to Schr\"odinger, Ludwig Boltzmann used large deviation type estimates in his pioneering work \citeTh[C]{Bol1877} connecting thermodynamics with the probability calculus.
The rigorous theory of large deviations started perhaps seven years after Schr\"odinger's paper in 1938 with the work of Harald Cram\'er \citeTh[C]{Cra1938} motivated by the ruin problem in insurance mathematics.  In the modern literature, an estimate like (\ref{me:ld}) is usually referred to as a ``level 2'' large deviation whose precise mathematical formulation goes under the name of Sanov's lemma \citeTh[C]{San1957}. 

We refer to \citeTh[C]{Tou2009} or to chapter~6 of \citeTh[C]{PePi2020} for an overview of large deviation theory aimed at a physics readership.  More mathematically oriented  references on modern large deviation theory are e.g. \citeTh[C]{Ell2005,DeZe2009}.

\subsection{Optimization problem in the continuum: ``static'' Schr\"odinger's problem}
\label{sec:Ssvp}

We are now ready to reformulate the problem in a formal continuum limit. 
As our aim is to emphasize the connection with Landauer's bound and related contemporary problems in statistical physics (see section~\ref{sec:Landauer} below), we consider a straightforward generalization of the continuum limit by replacing the sum over indices in  (\ref{me:re}) with integrals over~$\mathbb{R}^{d} $.  
The counterpart of the normalization condition (\ref{me:normalization}) is then
\begin{align}
	\nonumber
	\int_{\mathbb{R}^{2\,d}}\prod_{i=\mathfrak{0}}^{\mathfrak{1}}
	\mathrm{d}^{d}\bm{x}_{i}\,k(\bm{x}_{\mathfrak{1}},\bm{x}_{\mathfrak{0}})=1.
\end{align}
The continuum limit conditional probability is
\begin{align}
	\nonumber
	\mathscr{k}(\bm{x}_{\mathfrak{1}}|\bm{x}_{\mathfrak{0}})
	=	
	\frac{k(\bm{x}_{\mathfrak{1}},\bm{x}_{\mathfrak{0}})} {\,w_{\mathfrak{0}}(\bm{x}_{\mathfrak{0}})}.
\end{align}
The probability density $ w_{\mathfrak{0}} $ is the generalization over $\mathbb{R}^{d}  $ of the same quantity for $ d=1 $ considered by Schr\"odinger.

Next, we fix a reference transition probability density $\mathscr{g}$.  The counterpart of  the problem posed by Schr\"odinger reads as follows: \emph{finding the transition probability $\mathscr{k}$ minimizing its Kullback--Leibler 
divergence from $\mathscr{g}$ under the constraint that $\mathscr{k}$ evolves an initial density
$w_{\mathfrak{0}}$ into a final density $w_{\mathfrak{1}}$}. Mathematically, this is equivalent to find  $ \mathscr{k} $ as the minimizer of the functional  
\begin{align}
	\label{me:opt}
	\lefteqn{
		\mathcal{A}(\mathscr{k},\lambda_{\mathfrak{0}},\lambda_{\mathfrak{1}})=\int_{\mathbb{R}^{2d}}\prod_{i=\mathfrak{0}}^{\mathfrak{1}}
		\mathrm{d}^{d}\bm{x}_{i}\,\mathscr{k}(\bm{x}_{\mathfrak{1}}|\bm{x}_{\mathfrak{0}})w_{\mathfrak{0}}(\bm{x}_{\mathfrak{0}})
		\ln \frac{\mathscr{k}(\bm{x}_{\mathfrak{1}}|\bm{x}_{\mathfrak{0}})}{\mathscr{g}(\bm{x}_{\mathfrak{1}}|\bm{x}_{\mathfrak{0}})}
	}
	\nonumber\\&
	+\int_{\mathbb{R}^{2\,d}} \prod_{i=\mathfrak{0}}^{\mathfrak{1}}
	\mathrm{d}^{d}\bm{x}_{i}\,\Big{(}
	\lambda_{\mathfrak{1}}(\bm{x}_{\mathfrak{1}})
	+\lambda_{\mathfrak{0}}(\bm{x}_{\mathfrak{0}})\Big{)}\Big{(}w_{\mathfrak{1}}(\bm{x}_{\mathfrak{1}})
	-\mathscr{k}(\bm{x}_{\mathfrak{1}}|\bm{x}_{\mathfrak{0}})\Big{)}\,w_{\mathfrak{0}}(\bm{x}_{\mathfrak{0}}
	)
\end{align}
for $ w_{i} $, $ i=\mathfrak{0},\mathfrak{1} $ and $ \mathscr{g} $ given.
Of the integrals appearing in $  \mathcal{A}$
\begin{itemize}
	\item the  integral  appearing in the first line of (\ref{me:opt}) is the  Kullback--Leibler divergence between $ \mathscr{k} $ and $ \mathscr{g} $ \emph{averaged} with respect to the initial density $ w_{\mathfrak{0}} $. When dealing with the Kullback--Leibler divergence between transition probabilities of  two Markov processes, we always imply here also averaging with respect to the initial density.
	\item The integral having as a prefactor the \emph{Lagrange multiplier} $\lambda_{\mathfrak{1}}$ enforces  $\mathscr{k}$ to
	map the assigned initial density $w_{\mathfrak{0}}$ into $w_{\mathfrak{1}}$;
	\item  the integral associated to the \emph{Lagrange multiplier} $\lambda_{\mathfrak{0}}$ enforces $\mathscr{k}$ to preserve probability.
\end{itemize}
In what follows we denote by $ \mathscr{k}^{\prime} $ an arbitrary variation of $ \mathscr{k} $ satisfying the constraint
\begin{align}
	\nonumber
	\int_{\mathbb{R}^{d}}\mathrm{d}^{d}\bm{x}_{\mathfrak{2}}\,\mathscr{k}^{\prime}(\bm{x}_{\mathfrak{2}}|\bm{x}_{\mathfrak{1}}) = 0.
\end{align}
The stationary variation 
\begin{align}
\left .	\frac{\mathrm{d}}{\mathrm{d}\varepsilon }\right |_{\varepsilon=0}\mathcal{A}(\mathscr{k}+\varepsilon\,\mathscr{k}^{\prime},\lambda_{\mathfrak{0}},\lambda_{\mathfrak{1}})=0
	\nonumber
\end{align}
yields the condition
\begin{align}
	\nonumber
	w_{\mathfrak{0}}(\bm{x}_{\mathfrak{0}})
	\ln \frac{\mathscr{k}(\bm{x}_{\mathfrak{1}}|\bm{x}_{\mathfrak{0}})}{\mathscr{g}(\bm{x}_{\mathfrak{1}}|\bm{x}_{\mathfrak{0}})}
	-\lambda_{\mathfrak{1}}(\bm{x}_{\mathfrak{1}})\,w_{\mathfrak{0}}(\bm{x}_{\mathfrak{0}})-\lambda_{\mathfrak{0}}(\bm{x}_{\mathfrak{0}})\,w_{\mathfrak{0}}(\bm{x}_{\mathfrak{0}})=0
\end{align}
with solution $\mathscr{k}(\bm{x}_{\mathfrak{1}}|\bm{x}_{\mathfrak{0}})= \mathscr{g}(\bm{x}_{\mathfrak{1}}|\bm{x}_{\mathfrak{0}}) \,e^{\lambda_{\mathfrak{1}}(\bm{x}_{\mathfrak{1}})+\lambda_{\mathfrak{0}}(\bm{x}_{\mathfrak{0}})}$.
Similarly, arbitrary variations of $ \mathcal{A} $ with respect to the Lagrange multipliers yield the self-consistency conditions
\begin{align}
	\label{me:eq1}
	w_{\mathfrak{1}}(\bm{x}_{\mathfrak{1}})
	&=
	\int_{\mathbb{R}^{d}}\mathrm{d}^{d}\bm{x}_{\mathfrak{0}}\,e^{\lambda_{\mathfrak{1}}(\bm{x}_{\mathfrak{1}})}
	\mathscr{g}(\bm{x}_{\mathfrak{1}}|\bm{x}_{\mathfrak{0}})\,
	e^{\lambda_{\mathfrak{0}}(\bm{x}_{\mathfrak{0}})}\,w_{\mathfrak{0}}(\bm{x}_{\mathfrak{0}})
	\\
	\intertext{and}
	\label{me:eq2}
	1
	&=
	\int_{\mathbb{R}^{d}}
	\mathrm{d}^{d}\bm{x}_{\mathfrak{1}}\,e^{\lambda_{\mathfrak{1}}(\bm{x}_{\mathfrak{1}})}
	\mathscr{g}(\bm{x}_{\mathfrak{1}}|\bm{x}_{\mathfrak{0}})\,e^{\lambda_{\mathfrak{0}}(\bm{x}_{\mathfrak{0}})}.
\end{align}
Finally, upon setting $\varphi_{\mathfrak{1}}(\bm{x}_{\mathfrak{1}})=e^{\lambda_{\mathfrak{1}}(\bm{x}_{\mathfrak{1}})}$ and 
$w_{\mathfrak{0}}(\bm{x}_{\mathfrak{0}})=\varphi_{\mathfrak{0}}(\bm{x}_{\mathfrak{0}})\,e^{-\lambda_{\mathfrak{0}}(\bm{x}_{\mathfrak{0}})}$
we impose the boundary conditions in the form of Schr\"odinger's \emph{mass transport} equations (\ref{13'}) 
\begin{subequations}
	\label{me:sy}
	\begin{align}
		\label{me:sy1}
		w_{\mathfrak{1}}(\bm{x}_{\mathfrak{1}})=\varphi_{\mathfrak{1}}(\bm{x}_{\mathfrak{1}})
		\int_{\mathbb{R}^{d}}\mathrm{d}^{d}\bm{x}_{\mathfrak{0}}\,
		\mathscr{g}(\bm{x}_{\mathfrak{1}}|\bm{x}_{\mathfrak{0}})\,
		\varphi_{\mathfrak{0}}(\bm{x}_{\mathfrak{0}}),
\\
		\label{me:sy2}
		w_{\mathfrak{0}}(\bm{x}_{\mathfrak{0}})=\varphi_{\mathfrak{0}}(\bm{x}_{\mathfrak{0}})
		\int_{\mathbb{R}^{d}}\mathrm{d}^{d}\bm{x}_{\mathfrak{1}}\,\varphi_{\mathfrak{1}}(\bm{x}_{\mathfrak{1}})
		\mathscr{g}(\bm{x}_{\mathfrak{1}}|\bm{x}_{\mathfrak{0}}).
	\end{align}
\end{subequations}
The existence and uniqueness of the pair $ \varphi_{\mathfrak{0}}$, $\varphi_{\mathfrak{1}} $ solving (\ref{me:sy}) was later proven by Robert Fortet \citeTh[C]{For1940} and under weaker hypotheses by Arne Beurling \citeTh[C]{Beu1960} and Benton Jamison \citeTh[C]{Jam1974}. Further notable extensions and refinements have then been considered in \citeTh[C]{Foe1988,Nag1989,AeNa1992,Nag1993, Aeb1995}.

\subsection{Probability at intermediate times}
\label{sec:it}

We now start making explicit use of the Markov property. We identify the reference transition probability density $ \mathscr{g} $ with the value at $ t=t_{\mathfrak{1}} $, $ s=t_{\mathfrak{0}} $ of a two-parameter family of Markov transition probabilities $  \mathscr{g}_{t\,s} $.  
For any $ t $ belonging to the closed interval $ [t_{\mathfrak{0}}\,,t_{\mathfrak{1}}] $, we then construct a probability density $  w_{t}$  according to the formula 
(equation (\ref{14}) of Schrodinger)
\begin{align}
	w_{t}(\bm{x})= \bar{h}_{t}(\bm{x})\,h_{t}(\bm{x})
	\label{it:wt}
\end{align} 
with
\begin{subequations}
	\label{it:Born}
	\begin{align}
		&\label{it:Bornf}
		\bar{h}_{t}(\bm{x})=\int_{\mathbb{R}^{d}}\mathrm{d}^{d}\bm{x}_{\mathfrak{0}}\,\mathscr{g}_{t\,t_{\mathfrak{0}}}(\bm{x}|\bm{x}_{\mathfrak{0}})	\varphi_{\mathfrak{0}}(\bm{x}_{\mathfrak{0}}),
		\\
		&\label{it:Bornb}
			h_{t}(\bm{x})=\int_{\mathbb{R}^{d}}\mathrm{d}^{d}\bm{x}_{\mathfrak{1}}\,\varphi_{\mathfrak{1}}(\bm{x}_{\mathfrak{1}})
			\mathscr{g}_{t_{\mathfrak{1}}\,t}(\bm{x}_{\mathfrak{1}}|\bm{x}).
	\end{align}
\end{subequations}
The probability density (\ref{it:wt}) satisfies the conditions (\ref{me:sy})  as a consequence of the fact that the transition probability of a Markov process reduces in the limit $ |t-s|\downarrow 0 $ to the kernel of the identity operator
\begin{align}
	\lim_{|t-s|\downarrow 0}\mathscr{g}_{t\,s}(\bm{x}|\bm{y})=\delta^{(d)}(\bm{x}-\bm{y}).
	\nonumber
\end{align}
Overall (\ref{it:wt}) is  the source of what Schr\"odinger calls \emph{``remarkable analogies with quantum mechanics that seem to me worth considering''}. Namely (\ref{it:wt}) bears a formal resemblance with Born's rule prescribing that probabilities in Quantum Mechanics must be computed as the modulus squared of a probability amplitude.  
Furthermore, in Quantum Mechanics, the complex conjugation operation is interpreted as time reversal.  
Schr\"odinger's quote of Eddington's Gifford lecture remark in \S~4 of the paper refers to this fact.  
In the case of (\ref{it:wt}) time reversal is encoded in the definitions  (\ref{it:Bornf}), (\ref{it:Bornb}) respectively  stating that $  \bar{h}_{t}$ evolves as a particular solution of  Kolmogorov's forward equation admitting $ \mathscr{g}_{t\,t_{\mathfrak{0}}} $ as fundamental solution and that $ h_{t} $ is a harmonic function with respect to $\mathscr{g}_{t_{\mathfrak{1}\,t}}$:  a particular solution of Kolmogorov's backward equation also admitting  $\mathscr{g}_{t_{\mathfrak{1}}\,t}$ as fundamental solution; see e.g.~\citeTh[C]{Nag1993,Aeb1996,Pav2014} for further details.

We now want to show how  to directly obtain (\ref{it:wt}) as the solution of a dynamical optimal control problem \citeTh[C]{DaPr1991}.

\section{Stochastic optimal control problem}

\subsection{Relation with the pathwise Kullback-Leibler entropy minimization: 
	``dynamic'' Schr\"odinger diffusion problem}
\label{sec:KL}

We recall that the transition probability of a Markov process $ \mathscr{k}_{t\,s} $ obeys for any $s\,\leq\,u\,\leq\,t$ the Chapman-Kolmogorov equation (see e.g. \citeTh[C]{Pav2014}):
\begin{align}
	\mathscr{k}_{t\,s}(\bm{x}|\bm{y})=\int_{\mathbb{R}^{d}}\mathrm{d}\bm{z}\,\mathscr{k}_{t\,u}(\bm{x}|\bm{z})\mathscr{k}_{u\,s}(\bm{z}|\bm{y}).
	\label{KL:CK}
\end{align}
The  Kullback--Leibler  between between transition probability  $ \mathscr{k} $ and reference  transition probability  $ \mathscr{g} $ is
\begin{align}
	\nonumber
	\mathrm{D}_{\scriptscriptstyle{KL}}(\mathscr{k}\|\mathscr{g})=\int_{\mathbb{R}^{2\,d}}\prod_{i=\mathfrak{0}}^{\mathfrak{1}}
	\mathrm{d}^{d}\bm{x}_{i}\,
	\ln \frac{\mathscr{k}_{\tf\,\ti}(\bm{x}_{\mathfrak{1}}|\bm{x}_{\mathfrak{0}})}{\mathscr{g}_{\tf\,\ti}(\bm{x}_{\mathfrak{1}}|\bm{x}_{\mathfrak{0}})}
	\,\mathscr{k}_{\tf\,\ti}(\bm{x}_{\mathfrak{1}}|\bm{x}_{\mathfrak{0}})w_{\mathfrak{0}}(\bm{x}_{\mathfrak{0}}).
\end{align} 
We notice that the Chapman-Kolmogorov equation  allows us to pick an arbitrary $ s_{\mathfrak{0}} $ such that  $ t_{\mathfrak{0}}\,\leq\,s_{\mathfrak{0}} \,\leq\,t_{\mathfrak{1}}$, and couch the Kullback--Leibler divergence into the form
\begin{gather}
	\label{KL:split}
	\begin{aligned}[b]
		&	\mathrm{D}_{\scriptscriptstyle{KL}}(\mathscr{k}\|\mathscr{g})=
		\int_{\mathbb{R}^{3\,d}}\prod_{i=\mathfrak{0}}^{\mathfrak{1}}
		\mathrm{d}^{d}\bm{x}_{i}\,\mathrm{d}^{d}\bm{y}_{\mathfrak{0}}
		\ln R_{\tf\,\mathfrak{s}_{0}\,\ti}(\bm{x}_{\mathfrak{1}},\bm{y}_{\mathfrak{0}},\bm{x}_{\mathfrak{0}})\,
		\,\mathscr{k}_{\tf\,s_{\mathfrak{0}}}(\bm{x}_{\mathfrak{1}}\,|\,\bm{y}_{\mathfrak{0}})
		\,\mathscr{k}_{s_{\mathfrak{0}}\,\ti}(\bm{y}_{\mathfrak{0}}\,|\,\bm{x}_{\mathfrak{0}})
		\,w_{\mathfrak{0}}(\bm{x}_{\mathfrak{0}})
		\\
		& +
		\int_{\mathbb{R}^{3\,d}}\prod_{i=\mathfrak{0}}^{\mathfrak{1}}
		\mathrm{d}^{d}\bm{x}_{i}\,\mathrm{d}^{d}\bm{y}_{\mathfrak{0}}
		\ln \frac{\mathscr{k}_{\tf\,s_{\mathfrak{0}}}(\bm{x}_{\mathfrak{1}}\,|\,\bm{y}_{\mathfrak{0}})
			\,\mathscr{k}_{s_{\mathfrak{0}}\,\ti}(\bm{y}_{\mathfrak{0}}\,|\,\bm{x}_{\mathfrak{0}})}{\mathscr{g}_{\tf\,s_{\mathfrak{0}}}(\bm{x}_{\mathfrak{1}}\,|\,\bm{y}_{\mathfrak{0}})
			\,\mathscr{g}_{s_{\mathfrak{0}}\,\ti}(\bm{y}_{\mathfrak{0}}\,|\,\bm{x}_{\mathfrak{0}})}\,\mathscr{k}_{\tf\,s_{\mathfrak{0}}}(\bm{x}_{\mathfrak{1}}\,|\,\bm{y}_{\mathfrak{0}})
		\,\mathscr{k}_{s_{\mathfrak{0}}\,\ti}(\bm{y}_{\mathfrak{0}}\,|\,\bm{x}_{\mathfrak{0}})
		\,w_{\mathfrak{0}}(\bm{x}_{\mathfrak{0}}),
	\end{aligned}
	\intertext{where}
	\nonumber
	R_{\tf\,\mathfrak{s}_{0}\,\ti}(\bm{x}_{\mathfrak{1}},\bm{y}_{\mathfrak{0}},\bm{x}_{\mathfrak{0}})
	= 
	\frac{\mathscr{k}_{\tf\,\ti}(\bm{x}_{\mathfrak{1}}\,|\,\bm{x}_{\mathfrak{0}})}{\mathscr{g}_{\tf\,\ti}(\bm{x}_{\mathfrak{1}}\,|\,\bm{x}_{\mathfrak{0}})}
	\frac{\mathscr{g}_{\tf\,s_{\mathfrak{0}}}(\bm{x}_{\mathfrak{1}}\,|\,\bm{y}_{\mathfrak{0}})}{\mathscr{k}_{\tf\,s_{\mathfrak{0}}}(\bm{x}_{\mathfrak{1}}\,|\,\bm{y}_{\mathfrak{0}})}
	\frac{\mathscr{g}_{s_{\mathfrak{0}}\,\ti}(\bm{y}_{\mathfrak{0}}\,|\,\bm{x}_{\mathfrak{0}})}{\mathscr{k}_{s_{\mathfrak{0}}\,\ti}(\bm{y}_{\mathfrak{0}}\,|\,\bm{x}_{\mathfrak{0}})}.
\end{gather}
The observation is useful if we then apply the inequality (\ref{me:ineq}) to the first integral on the right-hand side of (\ref{KL:split}). After straightforward algebra we obtain
\begin{gather}
	\nonumber
	\begin{split}
		&	\mathrm{D}_{\scriptscriptstyle{KL}}(\mathscr{k}\|\mathscr{g})
		\leq
		\int_{\mathbb{R}^{2\,d}}
		\mathrm{d}^{d}\bm{x}_{1}\,\mathrm{d}^{d}\bm{y}_{\mathfrak{0}}
		\ln \frac{\mathscr{k}_{\tf\,s_{\mathfrak{0}}}(\bm{x}_{\mathfrak{1}}\,|\,\bm{y}_{\mathfrak{0}})
		}{\mathscr{g}_{\tf\,s_{\mathfrak{0}}}(\bm{x}_{\mathfrak{1}}\,|\,\bm{y}_{\mathfrak{0}})
		}
		\, \mathscr{k}_{\tf\,s_{\mathfrak{0}}}(\bm{x}_{\mathfrak{1}}\,|\,\bm{y}_{\mathfrak{0}})
		\, w_{s_{\mathfrak{0}}}(\bm{y}_{\mathfrak{0}})
		\\
		&\hspace{0.5cm}	+\int_{\mathbb{R}^{2\,d}}\mathrm{d}^{d}\bm{y}_{\mathfrak{0}}
		\,\mathrm{d}^{d}\bm{x}_{\mathfrak{0}}
		\ln \frac{
			\,\mathscr{k}_{s_{\mathfrak{0}}\,\ti}(\bm{y}_{\mathfrak{0}}\,|\,\bm{x}_{\mathfrak{0}})}{
			\,\mathscr{g}_{s_{\mathfrak{0}}\,\ti}(\bm{y}_{\mathfrak{0}}\,|\,\bm{x}_{\mathfrak{0}})}\,
		\,\mathscr{k}_{s_{\mathfrak{0}}\,\ti}(\bm{y}_{\mathfrak{0}}\,|\,\bm{x}_{\mathfrak{0}})
		\,w_{\mathfrak{0}}(\bm{x}_{\mathfrak{0}}),
	\end{split}
	\intertext{where now}
	\label{KL:pdf}
	w_{t}(\bm{x})
	=
	\int_{\mathbb{R}^{d}}\mathrm{d}^{d}\bm{x}_{\mathfrak{o}}\mathscr{k}_{t\,\ti}(\bm{x}\,|\,\bm{x}_{\mathfrak{0}})
	\,w_{\mathfrak{0}}(\bm{x}_{\mathfrak{0}}).
\end{gather}
If we repeat the same steps over an arbitrary partition in $ n+2\,\geq\,3 $ sub-intervals of the time interval $ [\ti\,,\tf] $ we obtain
\begin{align}
	&	\mathrm{D}_{\scriptscriptstyle{KL}}(\mathscr{k}\|\mathscr{g})\leq \int_{\mathbb{R}^{2\,d}}\mathrm{d}^{d}\bm{y}_{\mathfrak{n}}
	\mathrm{d}^{d}\bm{x}_{\mathfrak{1}}\,\ln \frac{\mathscr{k}_{\tf\,s_{\mathfrak{n}}}(\bm{x}_{\mathfrak{1}}\,|\,\bm{y}_{\mathfrak{n}})
	}{\mathscr{g}_{\tf\,s_{\mathfrak{n}}}(\bm{x}_{\mathfrak{1}}\,|\,\bm{y}_{\mathfrak{n}})
	}\,\mathscr{k}_{\tf\,s_{\mathfrak{n}}}(\bm{x}_{\mathfrak{1}}\,|\,\bm{y}_{\mathfrak{n}})
	\,w_{s_{\mathfrak{n}}}(\bm{y}_{\mathfrak{n}})
	\nonumber\\
	&\hspace{0.5cm}+\sum_{i=0}^{n-1}\int_{\mathbb{R}^{2d}}\mathrm{d}^{d}\bm{y}_{\mathfrak{i}}\mathrm{d}^{d}\bm{y}_{\mathfrak{i+1}}
	\,\ln \frac{\mathscr{k}_{s_{\mathfrak{i+1}}\,s_{\mathfrak{i}}}(\bm{y}_{\mathfrak{i+1}}\,|\,\bm{y}_{\mathfrak{i}})
	}{\mathscr{g}_{s_{\mathfrak{i+1}}\,s_{\mathfrak{i}}}(\bm{y}_{\mathfrak{i+1}}\,|\,\bm{y}_{\mathfrak{i}})
	}\,\mathscr{k}_{s_{\mathfrak{i+1}}\,s_{\mathfrak{i}}}(\bm{y}_{\mathfrak{i+1}}\,|\,\bm{y}_{\mathfrak{i}})
	\,w_{s_{\mathfrak{i}}}(\bm{y}_{\mathfrak{i}})
	\nonumber\\
	&\hspace{0.5cm}+\int_{\mathbb{R}^{2d}}\mathrm{d}^{d}\bm{y}_{\mathfrak{0}}\mathrm{d}^{d}\bm{x}_{\mathfrak{0}}
	\ln \frac{\mathscr{k}_{s_{\mathfrak{0}}\,t_{\mathfrak{0}}}(\bm{y}_{\mathfrak{0}}\,|\,\bm{x}_{\mathfrak{0}})
	}{\mathscr{g}_{s_{\mathfrak{0}}\,t_{\mathfrak{0}}}(\bm{y}_{\mathfrak{0}}\,|\,\bm{x}_{\mathfrak{0}})
	}\,\mathscr{k}_{s_{\mathfrak{0}}\,t_{\mathfrak{0}}}(\bm{y}_{\mathfrak{0}}\,|\,\bm{x}_{\mathfrak{0}})
	\,w_{\mathfrak{0}}(\bm{x}_{\mathfrak{0}})
	\label{KL:iteration}
\end{align}
Passing to the limit $ n\uparrow\infty $ we may formally write 
\begin{align}
	\mathrm{D}_{\scriptscriptstyle{KL}}(\mathscr{k}\|\mathscr{g})
	\leq
	\mathrm{D}_{\scriptscriptstyle{KL}}(\mathrm{P}_{\mathscr{k}}\|\mathrm{P}_{\mathscr{g}}).
	\label{KL:ineq}
\end{align}
The right-hand side, if it exists, is the Kullback--Leibler divergence between the probability measure $ \mathrm{P}_{\mathscr{k}} $ generated by the Markov process with transition probability $ \mathscr{k} $ and initial density $w_{\mathfrak{0}}$  and the probability measure $ \mathrm{P}_{\mathscr{g}} $ of the reference process with transition probability $ \mathscr{g} $ and initial density $w_{\mathfrak{0}}$.  We call $  \mathrm{D}_{\scriptscriptstyle{KL}}(\mathrm{P}_{\mathscr{k}}\|\mathrm{P}_{\mathscr{g}})$ the \emph{pathwise Kullback--Leibler divergence}. 

From the physics side, the inequality (\ref{KL:ineq}) has a simple interpretation. The Kullback--Leibler divergence is a relative entropy. If we construe entropy as a quantity counting the relevant number of degrees of freedom,  physical intuition suggests that its value increases when measuring the divergence at each instant of time rather than once over the full time interval of the evolution.
Most importantly,  $  \mathrm{D}_{\scriptscriptstyle{KL}}(\mathrm{P}_{\mathscr{k}}\|\mathrm{P}_{\mathscr{g}})$ admits a direct expression in terms of  quantities characterizing the microscopic state of the Markov process  as we  turn to show in the following section.

\subsection{Explicit expression of  the pathwise Kullback--Leibler divergence from a microscopic dynamics}
\label{sec:sde}

From now on we set the focus on the pathwise Kullback--Leibler divergence $ \mathrm{D}_{\scriptscriptstyle{KL}}(\mathrm{P}_{\mathscr{k}}\|\mathrm{P}_{\mathscr{g}})$. Working with pathwise Kullback--Leibler divergence is natural for non-equilibrium statistical mechanics (\citeTh[C]{DaPr1991} and e.g.  \mbox{\citeTh[C]{Nag1993,LeSp1999,Sek2010,Gaw2013,PePi2020}}) and control theory (see e.g. \citeTh[C]{DaPrMeRu1996,DuEl1997,BiKa2014,ChTo2015,ThTo2012}) applications  precisely because of the existence of a direct link with the microscopic dynamics. The naturalness of this concept is the reason why  $ \mathrm{D}_{\scriptscriptstyle{KL}}(\mathrm{P}_{\mathscr{k}}\|\mathrm{P}_{\mathscr{g}})$ is often referred to in the statistical physics literature without the further specification of ``pathwise''.

In order to determine the explicit expression of the limit of (\ref{KL:iteration}) as the mesh of the partition of $ [\ti\,,\tf] $ goes to zero, we note that the  probability measures
of  $ \mathrm{P}_{\mathscr{k}} $ and $ \mathrm{P}_{\mathscr{g}} $ coincide with the path measures of  It\^o stochastic differential equations of the form
\begin{subequations}
	\label{sde:sde}
	\begin{align}
		&\label{sde:sde1}
		\mathrm{d}\bm{\xi}_{t}=\bigg{(}\bm{b}_{t}(\bm{\xi}_{t})+\bm{u}_{t}(\bm{\xi}_{t})\bigg{)}\mathrm{d}t
		+\mathsf{a}_{t}(\bm{\xi}_{t})\cdot\mathrm{d}\bm{\omega}_{t},
		\\
		&\label{sde:sde2}
		\mathrm{Pr}\left(\bm{x}\,\leq\,\bm{\xi}_{t_{\mathfrak{0}}}<\bm{x}+\mathrm{d}\bm{x}\right)=w_{\mathfrak{0}}(\bm{x})\mathrm{d}^{d}\bm{x}.
	\end{align}
\end{subequations}
where $ \left\{ \bm{\omega}_{t} \right\}_{t\,\in \,[\ti\,,\tf]} $ is a Wiener process. 
The drift in (\ref{sde:sde1}) is the sum of two vector fields $\bm{b}\,,\bm{u}\colon[\ti,\tf]\,\times\,\mathbb{R}^{d}\mapsto\mathbb{R}^{d}$, the second of which, $ \bm{u} $, called the \emph{control}, we take to be identically vanishing in the reference case.
The diffusion amplitude $\mathsf{a}$ is  a position-dependent \emph{strictly positive definite matrix} $\mathsf{a}_{t}\colon[\ti,\tf]\,\times\,\mathbb{R}^{d}\mapsto\mathbb{R}^{d^{2}}$ related to the diffusion matrix by $\operatorname{A}_{t}(\bm{x})=(\mathsf{a}_{t} \mathsf{a}_{t}^{\top})(\bm{x}).$
The It\^o differential equation (\ref{sde:sde}) representation of the dynamics provides an explicit expression of the transition probability in any infinitesimal interval belonging to a partition of $ [\ti\,,\tf] $ (see e.g. chapter~5 of \citeTh[C]{Sch2010}):
\begin{equation}
	\label{sde:st}
	\mathscr{k}_{t_{\mathfrak{i+1}}\,t_{\mathfrak{i}}}(\bm{x}_{\mathfrak{i+1}}\,|\,\bm{x}_{\mathfrak{i}})
	=
	\frac{
		\exp\left(
			-\frac{1}{2} \left\|
				\frac{\boldsymbol{x}_{i+1}-\boldsymbol{x}_{i}}{\tau_{ i}} 
				-(\boldsymbol{b}_{t_{i}}+\boldsymbol{u}_{t_{i}})(\boldsymbol{x}_{i})
			\right\|_{\mathsf{A}_{t_{i}}^{-1}(\boldsymbol{x}_{i})}^{2}\,\tau_{i}
		\right)
	}{(
		2\,\pi\,\det\mathsf{A}_{t_{i}}(\bm{x}_{i})\,\tau_{i})^{d/2}
	} + o(\tau_{i}).
\end{equation}
Here we use the notation $\|\bm{v}\|_{\mathsf{A}^{-1}}^{2}=\langle \bm{v}\,,\mathsf{A}^{-1}\bm{v}\rangle_{\mathbb{R}^{d}}$ relating the squared norm with metric $ \operatorname{A}^{-1} $ of a vector to the inner product in $ \mathbb{R}^{d} $, and $\tau_{i}=t_{i+1}-t_{i}$.
The short-time expression of the transition probability immediately implies
\begin{align}
	\nonumber
	\lefteqn{
		\ln\frac{\mathscr{k}_{t_{i+1}\,t_{i}}(\bm{x}_{i+1}|\bm{x}_{i})}
		{\mathscr{g}_{t_{i+1}\,t_{i}}(\bm{x}_{i+1}|\bm{x}_{i})}
		=
	}
	\nonumber
	\\
	&\qquad
	\left(\frac{\langle\bm{x}_{i+1}-\bm{x}_{i}-\bm{b}_{t_{i}}(\bm{x}_{i})\,,
		(\mathsf{A}_{t_{i}}^{-1}\bm{u}_{t_{i}})(\bm{x}_{i})\rangle_{\mathbb{R}^{d}}-\left\|\bm{u}_{t_{i}}(\bm{x}_{i})\right\|_{\mathsf{A}_{t_{i}}^{-1}(\bm{x}_{i})}^{2}}{2}\right )\tau_{i}
	+ o(\tau_{i}).
	\nonumber
\end{align}
On each sub-interval, the integral over the variable $\bm{x}_{i+1}$ is Gaussian and after an obvious change of  variables becomes
\begin{align}
	\nonumber
		\int_{\mathbb{R}^{d}}\mathrm{d}^{d}\bm{x}_{i+1}
		\,\mathscr{k}_{t_{i+1}\,t_{i}}(\bm{x}_{i+1}|\bm{x}_{i})
		\ln \frac{
			\,\mathscr{k}_{t_{i+1}\,t_{i}}(\bm{x}_{i+1}|\bm{x}_{i})}{
			\mathscr{g}_{t_{i+1}\,t_{i}}(\bm{x}_{i+1}|\bm{x}_{i})}
	=
	\frac{\tau_{ i}}{2}
	\left\|\bm{u}_{t_{i}}(\bm{x}_{i})\right\|_{\operatorname{A}_{t_{i}}^{-1}(\bm{x}_{i})}^{2}+o(\tau_{i}).
\end{align}
Passing to the limit we finally arrive at
\begin{align}
		D_{\scriptscriptstyle{KL}}(\mathrm{P}_{\mathscr{k}}\|\mathrm{P}_{\mathscr{g}})
	=\int_{t_{o}}^{t_{f}}\mathrm{d}t\,
	\int_{\mathbb{R}^{d}}\mathrm{d}^{d}\bm{x}\,
	\,\frac{\|\bm{u}_{t}(\bm{x})\|_{\mathsf{A}_{t}^{-1}(\bm{x})}^{2}}{2}\,
	w_{t}\left(\bm{x}\right)
	\label{sde:KL}
\end{align}
with $ w_{t}\left(\bm{x}\right)$ evolving  according to (\ref{KL:pdf}).

\textbf{Remark.}
	A further generalization is obtained if we take as reference process  the system of It\^o stochastic differential equations
		\begin{align}
			\nonumber
	&		\mathrm{d}\bm{\xi}_{t}=\bm{b}_{t}(\bm{\xi}_{t})\mathrm{d}t
			+\mathsf{a}_{t}(\bm{\xi}_{t})\mathrm{d}\bm{w}_{t},
\\
&			\mathrm{d}\zeta_{t}=-V_{t}(\bm{\xi}_{t})\zeta_{t}\mathrm{d}t,
\nonumber
		\end{align}
	where $V\colon\mathbb{R}^{d}\times[\ti\,,\tf]\mapsto\mathbb{R}_{+}$. The reference pseudo-transition probability density admits the short-time representation
	\begin{align}
			\mathscr{g}_{t_{i+1}\,t_{i}}(\bm{x}_{i+1}|\bm{x}_{i})
			=
		\frac{\exp\left(-\frac{\tau_{i}}{2}
			\left\|\frac{\bm{x}_{i+1}-\bm{x}_{i}}{\tau_{i}}
			-\bm{b}_{t_{i}}(\bm{x}_{i})
			\right\|_{\mathsf{A}_{t_{i}}^{-1}(\bm{x}_{i})}^{2}-V_{t_{i}}(\bm{x}_{i})\tau_{i}\right)}{(2\,\pi\,\det\mathsf{A}_{t_{i}}(\bm{x}_{i})\,\tau_{i})^{d/2}}+o(\tau_{i}).
		\nonumber
	\end{align}
	In such a case we obtain the functional
	\begin{align}
		\label{sde:potential}
		D_{\scriptscriptstyle{KL}}(\mathrm{P}_{\mathscr{k}}\|\mathrm{P}_{\mathscr{g}})
		=
		\int_{t_{o}}^{t_{f}}\mathrm{d}t\,\int_{\mathbb{R}^{2 d}}\mathrm{d}^{d}\bm{x}\,
		\left(
			\frac{\|\bm{u}_{t}(\bm{x})\|_{\mathsf{A}_{t}^{-1}(\bm{x})}^{2}}{2} +V_{t}(\bm{x})
		\right) \, w_{t}(\bm{x}).
	\end{align}
We refer to the stochastic mechanics literature (see e.g. \citeTh[C]{Nag1964,Nag1993,Aeb1996,Zam2009} and references therein) for a rigorous derivation and physical  interpretation of this result.	
	\begin{center}
		* *
	\end{center}

\subsection{Infinite dimensional optimal control problem}
\label{sec:vp}

\begin{figure}[H]
	\centering
	\begin{subfigure}[t]{0.5\textwidth}
		\centering
		\includegraphics[width=0.7\textwidth]{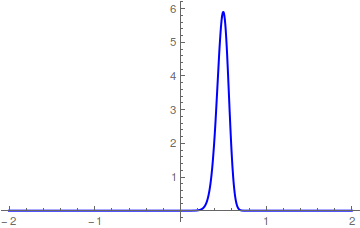}
		\caption{Initial state: probability peaked on one state }
		\label{fig:initial}
	\end{subfigure}%
	\begin{subfigure}[t]{0.5\textwidth}
		\centering
		\includegraphics[width=0.7\textwidth]{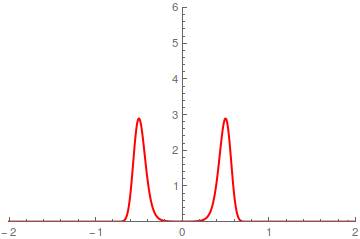}
		\caption{Final state: probability density equally peaked on two states }
		\label{fig:longtime}
	\end{subfigure}
	\caption{Pictorial description of a Schr\"odinger diffusion problem corresponding to the erasure of one bit of memory. According to Landauer's principle only logically irreversible operations are in principle thermodynamically irreversible i.e. correspond to dissipative processes. Since other logical operations can be implemented reversibly, erasure would therefore be the only irreversible operation in the thermodynamics of computation. The controlled Markovian dynamics embodying a physically meaningful realization of the principle corresponds then to the minimizer of an adapted thermodynamic quantity. We discuss in section~\ref{sec:Landauer} below the relation between the Kullback--Leibler divergence considered by Schr\"odinger and that entering the currently accepted formulation of the principle.}
	\label{fig:Landauer}
\end{figure}
\setlength{\unitlength}{1cm}
\begin{picture}(0,0)
	\put (7.5,7){\includegraphics[height=0.08\textwidth]{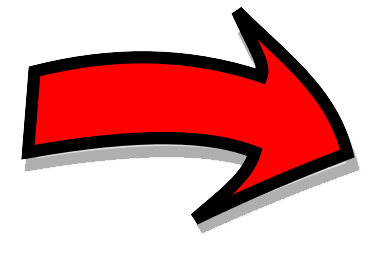}}
	\put(6,6.5){\textcolor{darkred}{Controlled Markovian dynamics}}
\end{picture}
The optimal control problem associated to (\ref{sde:KL}) is: \emph{once  the probability densities $ w_{\mathfrak{0}} $ and $ w_{\mathfrak{1}} $ are assigned at the boundaries of the control interval $ [\ti\,,\tf] $, find the vector field $ \bm{u} $ in (\ref{sde:sde1}) such that $ w_{\mathfrak{0}} $ evolves into $ w_{\mathfrak{1}} $ whilst minimizing the Kullback--Leibler divergence (\ref{sde:KL})}. To the best of our knowledge,  this formulation of Schr\"odinger's problem is due to Hans F\"ollmer \citeTh[C]{Foe1988}.

One way to derive the optimal control equations in analogy with what is done in the finite dimensional case  (\ref{me:opt}) is to apply the method of the so called ``adjoint equation'',
well known in statistical hydrodynamics \citeTh[C]{Ser1959,SeWh1968}.
The idea is to reformulate optimal control as a variational problem for an action functional whereby the dynamics are imposed by means of Lagrange multipliers. We may conceptualize the adjoint equation method as an extension of Pontryagin's principle (see e.g. \citeTh[C]{Lib2012}) to stochastic dynamics in analogy with Jean-Michel Bismut's treatment of stochastic variational calculus (\citeTh[C]{Bis1979} see also \citeTh[C]{KoPa1993}). We refer to \citeTh[C]{DaPr1991} for a mathematically rigorous treatment of the optimal control problem  whereas \citeTh[C]{Bec2021} provides an overview on optimal control in general and targeted at the physics audience.

\subsubsection{The adjoint equation formulation of optimal control}

In order to illustrate the adjoint equation method, we recall that the mean forward derivative of a test scalar function $ f $ along the paths of (\ref{sde:sde}) is 
\begin{align}
	\nonumber
	\operatorname{L}_{\bm{x}}f(\bm{x})
	&=
	\lim_{\varepsilon\downarrow 0}\operatorname{E}\left(
	\frac{f(\bm{\xi}_{t+\varepsilon})-f(\bm{\xi}_{t})}{\varepsilon}\bigg{|}\bm{\xi}_{t}=\bm{x}\right)
	\\
	\label{vp:mfd}
	&=
	\left \langle\,(\bm{b}_{t}+\bm{u}_{t})(\bm{x})\,,\partial_{\bm{x}}f(\bm{x})\,\right\rangle
	+\frac{1}{2}\operatorname{Tr}(\operatorname{A}_{t}(\bm{x})\partial_{\bm{x}}\otimes\partial_{\bm{x}})f(\bm{x}).
\end{align}
The mean forward derivative is thus specified by the action of a differential operator $  \operatorname{L}$, called the generator, on test scalar functions \citeTh[C]{Sch2010}. 
The expression of mean forward derivative along the paths of the reference process is readily seen by setting $ \bm{u}_{t}=0 $ in the foregoing definition (\ref{vp:mfd}).
The adjoint equation method consists of determining the optimal control equations by imposing that the functional
\begin{align}
	\MoveEqLeft
	\mathcal{A}[J,u,w]=
- \int_{\mathbb{R}^{d}}\mathrm{d}\bm{x}\bigg{(}w_{\mathfrak{1}}(\bm{x})\,J_{\tf}(\bm{x})-w_{\mathfrak{0}}(\bm{x})\,J_{\ti}(\bm{x})\bigg{)}
	\nonumber
	\\
	&+ \int_{\ti}^{\tf}\mathrm{d}t\,\int_{\mathbb{R}^{d}}\mathrm{d}^{d}\bm{x} \,w_{t}(\bm{x})\left(\frac{\|\bm{u}_{t}(\bm{x})\|_{\mathsf{A}_{t}^{-1}(\bm{x})}^{2}}{2}
	+(\partial_{t}+\operatorname{L}_{\bm{x}})J_{t}(\bm{x})
	\right)
	\label{vp:A}
\end{align}
be stationary with respect to the fields $ J $, $ w $ and $ \bm{u} $. To justify  (\ref{vp:A}) we observe that  the field $ J $ plays the role of a Lagrange multiplier
enforcing  the evolution law that the probability density $ w_{t} $ must obey in $ [\ti\,\tf] $. Namely, an integration by parts over  position variables defines the adjoint $ \operatorname{L}^{\dagger} $ of the generator $ \operatorname{L} $ 
\begin{align}
	\int_{\mathbb{R}^{d}}\mathrm{d}^{d}\bm{x}w_{t}(\bm{x})\operatorname{L}_{\bm{x}}J_{t}(\bm{x})=\int_{\mathbb{R}^{d}}\mathrm{d}^{d}\bm{x}J_{t}(\bm{x})\operatorname{L}_{\bm{x}}^{\dagger}w_{t}(\bm{x}).
	\nonumber
\end{align}
Similarly, an integration by parts with respect to the time variable brings about the identity
\begin{align}
	\int_{\ti}^{\tf}\mathrm{d}t\,\int_{\mathbb{R}^{d}}\mathrm{d}^{d}\bm{x}\,w_{t}(\bm{x})\partial_{t}J_{t}(\bm{x})=
	\int_{\mathbb{R}^{d}}\mathrm{d}\bm{x}\bigg{(}w_{\mathfrak{1}}(\bm{x})\,J_{\tf}(\bm{x})-w_{\mathfrak{0}}(\bm{x})\,J_{\ti}(\bm{x})\bigg{)}
	-\int_{\ti}^{\tf}\mathrm{d}t\,\int_{\mathbb{R}^{d}}\mathrm{d}^{d}\bm{x}\,J_{t}(\bm{x})\partial_{t}w_{t}(\bm{x}),
	\nonumber
\end{align}
which allows us to couch (\ref{vp:A}) into the equivalent form
\begin{align}
		\mathcal{A}[J,u,w]=\int_{\ti}^{\tf}\mathrm{d}t\,\int_{\mathbb{R}^{d}}\mathrm{d}^{d}\bm{x}
		\left( \frac{\|\bm{u}_{t}(\bm{x})\|_{\mathsf{A}_{t}^{-1}(\bm{x})}^{2}}{2}
		-J_{t}(\bm{x})(\partial_{t}-\operatorname{L}_{\bm{x}}^{\dagger})
		\right)\,w_{t}(\bm{x}).
	\label{vp:AA}
\end{align}
The role of $ J $ as Lagrange multiplier thus becomes manifest. At variance with the generator $ \operatorname{L} $,  the  adjoint $ \operatorname{L}^{\dagger} $ is not in general a differential operator as it instead depends upon the boundary conditions imposed on the stochastic process (see e.g. \citeTh[C]{Sch2010}). This is ultimately the general reason for preferring  (\ref{vp:A}) over (\ref{vp:AA}) in the formulation of the adjoint equation method. Here, however, we always consider probabilities decaying sufficiently rapidly at infinity in the Euclidean space $ \mathbb{R}^{d} $. As a consequence
we are entitled to identify $\operatorname{L}^{\dagger}  $ with the differential operator specifying the Fokker-Planck equation governing the evolution of $ w_{t} $.

\subsubsection{Optimal control equations}

 The action functional (\ref{vp:A}) is stationary if the fields satisfy 
\begin{subequations}
	\label{vp:stat}
	\begin{align}
		&\label{vp:stat1}
		\partial_{t}w_{t}=\operatorname{L}_{\bm{x}}^{\dagger}w_{t}(\bm{x}),
		\\
		&\label{vp:stat2}
		(\partial_{t}+\operatorname{L}_{\bm{x}})J_{t}(\bm{x})=-\frac{\|\bm{u}_{t}(\bm{x})\|_{\mathsf{A}_{t}^{-1}(\bm{x})}^{2}}{2},
		\\
		&\label{vp:stat3}
		\operatorname{A}_{t}(\bm{x})^{-1}\bm{u}_{t}(\bm{x})+\partial_{\bm{x}}J_{t}(\bm{x}) = 0.
	\end{align}
\end{subequations}
We recognize that (\ref{vp:stat1}) is the Fokker-Planck equation and (\ref{vp:stat2}) the dynamic programming equation of control theory. The stationary condition occasions the identification of the Lagrange multiplier with the \emph{value function} of optimal control theory \citeTh[C]{Lib2012}. Finally equation (\ref{vp:stat3}) relates
the stationary value of the so far unknown vector field~$\bm{u}_{t}$ to the solution of the dynamic programming equation (\ref{vp:stat2}).

\subsection{Solution of the optimal control equations}
\label{sec:soce}

Once we insert (\ref{vp:stat3}) into (\ref{vp:stat2}) we obtain the Hamilton-Jacobi-Bellman equation
\begin{align}
	\label{vp:HJB}
	\partial_{t}J_{t}(\bm{x})+\left \langle\,\bm{b}_{t}(\bm{x})\,,\partial_{\bm{x}}J_{t}(\bm{x})\,\right\rangle
	+
	\frac{1}{2} \operatorname{Tr}\left( \operatorname{A}_{t}(\bm{x})\partial_{\bm{x}}\otimes\partial_{\bm{x} }\right) J_{t}(\bm{x}) - \frac{\|\partial_{\bm{x}}J_{t}(\bm{x})\|_{\mathsf{A}}^{2}}{2}
	= 0.
\end{align}
The logarithmic transform
\begin{align}
	\label{rediff:logtrf}
	J_{t}(\bm{x})=-\ln h_{t}(\bm{x})
\end{align}
then maps the Hamilton-Jacobi-Bellman (\ref{vp:HJB}) into a backward Kolmogorov equation with respect to the reference process  \citeTh[C]{DaPr1991}:
\begin{align}
	\label{rediff:h}
	\partial_{t}h_{t}(\bm{x})+\left \langle\,\bm{b}_{t}(\bm{x})\,,\partial_{\bm{x}}h_{t}(\bm{x})\,\right\rangle
	+
	\frac{1}{2} \operatorname{Tr}\left( \operatorname{A}_{t}(\bm{x})\partial_{\bm{x}}\otimes\partial_{\bm{x}} \right) h_{t}(\bm{x})
	= 0.
\end{align}
In other words the function $ h_{t}$ is $ \mathscr{g} $-harmonic. The knowledge of the $ \mathscr{g} $-harmonic function $ h_{t} $ allows us to  determine via (\ref{vp:stat3}) (\ref{rediff:logtrf}) the value of the optimal control:
\begin{align}
	\bm{u}_{t}(x)=\operatorname{A}_{t}(x)\partial_{x}\ln h_{t}(x).
	\nonumber
\end{align}
Next, a direct calculation proves that the transition probabilities of the optimal control and reference process are linked by Doob's transform of the transition probability \citeTh[C]{Doo1957}, which yields
\begin{align}
	\mathscr{k}_{t\,s}(\bm{x}|\bm{y})=
	\frac{\mathscr{g}_{t\,s}(\bm{x}|\bm{y})\,h_{t}(\bm{x})}{
		h_{s}(\bm{y})}.
		\label{rediff:tp}
\end{align}
The solution of the optimal control problem is thus fully specified if we determine the boundary conditions for  $ h_{t} $ from the solution of Schr\"odinger's mass transport problem (\ref{me:sy}), where we now write
\begin{align}
&	\varphi_{\mathfrak{0}}(\bm{x})=\frac{w_{\mathfrak{0}}(\bm{x})}{h_{\ti}(\bm{x})},
\nonumber	\\
&	\varphi_{\mathfrak{1}}(\bm{x})=h_{\tf}(\bm{x}).
	\nonumber
\end{align}
The definition of transition probability density implies that the probability density of the optimal control process  evolves from $ t_{\mathfrak{0}} $ as
\begin{align}
	w_{t}(\bm{x})=\int_{\mathbb{R}^{d}} \mathrm{d}\bm{x}_{0}\,\mathscr{k}_{t\,t_{\mathfrak{0}}}(\bm{x}|\bm{x}_{\mathfrak{0}})w_{\mathfrak{0}}(\bm{x}_{\mathfrak{0}}).
	\nonumber
\end{align}
Upon inserting (\ref{rediff:tp}) in the above expression we obtain
\begin{align}
		w_{t}(\bm{x})=h_{t}(\bm{x})\int_{\mathbb{R}^{d}} \mathrm{d}\bm{x}_{0}\,\mathscr{g}_{t\,t_{\mathfrak{0}}}(\bm{x}|\bm{x}_{\mathfrak{0}})\varphi_{\mathfrak{0}}(\bm{x}_{\mathfrak{0}})=h_{t}(\bm{x})\bar{h}_{t}(\bm{x}).
	\label{rediff:w}
\end{align}
We thus recover the Born -like representation (\ref{it:wt}) of the probability density. In particular, the function  $ \bar{h}_{t} $ defined in (\ref{it:Bornf}) satisfies by construction  the forward Kolmogorov equation with respect to the reference process
\begin{subequations}
	\label{rediff:fKe}
	\begin{align}
		&\label{rediff:fKe1}
			\partial_{t}\bar{h}_{t}(\bm{x})+\left \langle\,\partial_{\bm{x}}\,,\bar{h}_{t}(\bm{x})\,\bm{b}_{t}(\bm{x})\,\right\rangle
		-\frac{1}{2}\operatorname{Tr}\left(\partial_{\bm{x}}\otimes\partial_{\bm{x}} \operatorname{A}_{t}(\bm{x})\bar{h}_{t}(\bm{x})\right) 
		= 0,
		\\
		&\label{rediff:fKe2}
		\bar{h}_{t_{\mathfrak{0}}}(\bm{x})
		= \varphi_{\mathfrak{0}}(\bm{x}).
	\end{align}
\end{subequations}
In summary we have shown that the problem posed and, modulo technical refinements, solved by Schr\"odinger is the optimal control problem of finding the diffusion process interpolating between two target states whilst minimizing the Kullback--Leibler divergence from
a reference uncontrolled process. We refer to \citeTh[C]{DaPr1991} see also \citeTh[C]{Bla1992,Aeb1996,Mik2004,LeZa2010,Leo2012,Leo2014} for further mathematical details.

\subsection{Connection with the Schr\"odinger equation}

The factorization of the interpolating probability (\ref{rediff:w}) admits a suggestive rewriting which further exhibits formal analogies with quantum mechanics.
Namely, if we introduce the complex ``wave function''
\begin{align}
	\label{Sceq:wv}
	\psi_{t}(\bm{x})=\sqrt{h_{t}(\bm{x})\bar{h}_{t}(\bm{x})} \exp\left(\frac{\imath}{2}\ln \frac{h_{t}(\bm{x})}{\bar{h}_{t}(\bm{x})}\right)
\end{align}
then Born's rule takes the expression familiar in quantum mechanics:
\begin{align}
	w_{t}(\bm{x})=\left |\psi_{t}(\bm{x})\right |^{2}.
	\nonumber
\end{align}
We emphasize that in (\ref{Sceq:wv}) amplitude and phase factors are well defined as $ h_{t} $ and $ \bar{h}_{t} $ are positive definite. 
Furthermore, the result of a tedious but conceptually straightforward calculation using Kolmogorov's forward (\ref{rediff:fKe1}) and backward (\ref{rediff:h}) equations and $A_{t}(\bm{x})=\frac{\hbar}{m}$ shows that the wave function (\ref{Sceq:wv})  satisfies 
\begin{align}
	\imath\,\partial_{t}\psi_{t}(\bm{x})=\frac{\hbar}{2\,m}\left (-\imath\,\partial_{\bm{x}}+\frac{m}{\hbar}\bm{b}_{t}(\bm{x})\right )^{2}\psi_{t}(\bm{x})
	+\left(U_{t}(\bm{x})-\frac{m}{2\,\hbar}\left\|\bm{b}_{t}(\bm{x})\right\|^{2}-\frac{1-\imath}{2}\partial_{\bm{x}}\cdot\bm{b}_{t}(\bm{x})\right)\psi_{t}(\bm{x}),
	\label{Sceq:Se}
\end{align}
where
\begin{align}
	U_{t}(\bm{x})=\frac{\hbar}{m}\left(\left\|\partial_{\bm{x}}\ln |\psi_{t}(\bm{x})|\right\|^{2}+\Delta_{\bm{x}}\ln |\psi_{t}(\bm{x})|\right).
	\label{Sceq:Madelung}
\end{align}
We thus recognize that the equation governing the wave function evolution is a non-linear Schr\"odinger equation for a particle in an electromagnetic field. Furthermore, had we taken as starting point the optimal control problem~(\ref{sde:potential}) then the linear potential term $ V_{t}(\bm{x}) $ would appear on the right-hand side of (\ref{Sceq:Se}).
This latter observation is meant to emphasize that we can adapt the formulation of Schr\"odinger's mass transport to recover all terms entering the most general form of  Schr\"odinger's equation in Quantum Mechanics.
The deep discrepancies between  Schr\"odinger's mass transport problem and Quantum Mechanics are encapsulated in the non-linear ``Madelung-de Broglie'' non-linear potential (\ref{Sceq:Madelung}) \citeTh[C]{Mad1927,deBro1957}. A further discrepancy with ordinary Quantum Mechanics is the existence of a kinematic equation for the position process
which we can straightforwardly derive by inserting the optimal value of the control into (\ref{sde:sde1}):
\begin{align}
	\mathrm{d}\bm{\xi}_{t}=\left(\bm{b}_{t}(\bm{\xi}_{t})+\frac{\hbar}{m}\left(\operatorname{Re}\partial_{\bm{\xi}_{t}}\ln \psi_{t}+\operatorname{Im}\partial_{\bm{\xi}_{t}}\ln \bar{\psi}_{t}\right)\right)\mathrm{d}t
	+\sqrt{\frac{\hbar}{m}}\cdot\mathrm{d}\bm{\omega}_{t}
	\nonumber
\end{align}

Most of the physics literature inspired by Schr\"odinger's paper has investigated the analogies between classical stochastic processes and quantum mechanics. It is impossible to give a fair account of this literature within this short commentary. We therefore restrict ourselves to a few observations that, we hope, might serve as an invitation to the existing excellent literature. 

Whilst Schr\"odinger and, as mentioned in the introduction, F\"urth \citeTh[C]{Fur1933} (see also \citeTh[C]{PePMG2020}) uncover classical probabilistic analogues of quantum mechanics, the perspective of F\'enyes, \citeTh[C]{Fen1952} and Nelson \citeTh[C]{Nel1985, Nel2001} is somewhat reversed as they try to reformulate quantum mechanics as a classical probabilistic theory. 
The aim is to show, in the words of  F\'enyes \citeTh[C]{Fen1952}, that ``\emph{wave mechanics processes are special Markov processes}'' and that ``\emph{the problem of the `hidden parameters' can also be solved in quantum mechanics using the principle of causality}'' thus arriving at a ``\emph{statistical derivation of the Schr\"odinger equation}''. 
We refer to \citeTh[C]{GrHaTa1979} for criticism (see \citeTh[C]{BlGoSe1986} for a reply) and to \citeTh[C]{FaGrSiBrCa2006} (see also \citeTh[C]{BlCoZh1987,Nag1993,Aeb1996}) for a state-of-the-art overview of this ambitious and controversial program.

Rich in applications, especially in numerical simulations of field theories, are imaginary-time models of finite and infinite dimensional quantum mechanics, which can be also traced back to the ideas put forward by Schr\"odinger and F\"urth.  In addition to the applications in optimal stochastic control mentioned in the introduction, it is worth mentioning the stochastic quantization proposed by Giorgi Parisi and Yongshi Wu in \citeTh[C]{PaWu1981}.  Stochastic quantization regards Euclidean quantum field theory as the equilibrium limit in an extra fictitious time variable of a statistical system coupled to a thermal reservoir. We refer to \citeTh[C]{DaHu1987} for a self-contained presentation.

\section{Relation with Kolmogorov's time reversal}
\label{sec:tr}

We now turn to discuss the implications for Schr\"odinger's mass transport equations (\ref{me:sy})  of the  time reversal relations for Markov processes 
described by Kolmogorov in  \citeTh[C]{Kol1936,Kol1937}. Schr\"odinger's and Kolmogorov's work originated a rich literature investigating properties of Markov processes under time reversal (see e.g. \citeTh[C]{Nag1964,HaPa1986,ChWa2005}) and consequences for irreversibility in statistical and quantum physics (see e.g. \citeTh[C]{LeSp1999,JiQiQi2004,ChGa2008} and \citeTh[C]{JaMa2003} for a pedagogic introduction).  Without any pretense of completeness, we only highlight here some elementary facts.

\pagebreak[3]
\subsection{Time reversal for Markov transition probability densities}

We start by assuming that we are given  
\begin{itemize}
	\item the transition probability density $ \mathscr{g}_{t\,s} $ of a Markov process for any $ s\,\leq\,t\,\in\,[\ti\,,\tf] $;
	\item a particular expression of the probability density of the Markov process $ p_{t} $ evolving from e.g. $ p_{\ti} $ at time $ \ti $ and \emph{strictly positive} for all $ t\,\in\,[\ti\,,\tf] $.
\end{itemize}
This information allows us to write the joint probability density of the Markov process  at any times $ s $ and $ t $
\begin{align}
	\mathscr{c}_{t\,s}(\bm{x},\bm{y})=\mathscr{g}_{t\,s}(\bm{x}|\bm{y})p_{s}(\bm{y}).
	\nonumber
\end{align}
Drawing from \citeTh[C]{Kol1936}, we then use the joint probability to define a time reversed transition probability associated to the density $ g_{t} $
\begin{align}
	\nonumber
	\mathscr{g}_{s\,t}^{(r)}(\bm{y}|\bm{x})=\frac{\mathscr{c}_{t\,s}(\bm{x},\bm{y})}{p_{t}(\bm{x})}=\frac{\mathscr{g}_{t\,s}(\bm{x}|\bm{y})p_{s}(\bm{y})}{p_{t}(\bm{x})}
	\shortintertext{or, equivalently,}
	\label{tr:K}
	p_{t}(\bm{x})
	\,\mathscr{g}_{s\,t}^{(r)}(\bm{y}|\bm{x})
	=
	\mathscr{g}_{t\,s}(\bm{x}|\bm{y})
	p_{s}(\bm{y}).
\end{align}
The time reversed transition probability density $ \mathscr{g}_{s\,t}^{(r)}(\bm{y}|\bm{x}) $ has  the interpretation of specifying the probability density of  the event that the process visits the state $ \bm{y} $ at a \emph{previous} time $ s $ \emph{conditional} upon the fact that we know that the process is in $ \bm{x} $ at a \emph{subsequent} time $ t $. 
A direct calculation (see e.g. \citeTh[C]{Nag1993} and also \citeTh[C]{Dyn1971,Dyn1978,ChWa2005,ChGu2011}) shows that if $ \mathscr{g}_{t\,s}  $ and $ p_{s} $ obey the same microscopic dynamics (\ref{sde:sde1}) then  $ \mathscr{g}_{s\,t}^{(r)} $ satisfies a pair of adjoint Kolmogorov equations 
\begin{subequations}
	\label{tr:pair}
	\begin{align}
		&\label{tr:pairf}
		\partial_{s} \mathscr{g}_{s\,t}^{(r)} (\bm{y}|\bm{x})+\left \langle\,\partial_{\bm{y}}\,,\mathscr{g}_{s\,t}^{(r)} (\bm{y}|\bm{x})\,\bm{b}_{s}^{(r)}(\bm{y})\,\right\rangle
		+\frac{1}{2}\operatorname{Tr}\left(\partial_{\bm{y}}\otimes\partial_{\bm{x}} \operatorname{A}_{s}(\bm{y})\mathscr{g}_{s\,t}^{(r)} (\bm{y}|\bm{x})\right) =
		0,
		\\
		&\label{tr:pairb}
		\partial_{t}	\mathscr{g}_{s\,t}^{(r)} (\bm{y}|\bm{x})+\left \langle\,\bm{b}_{t}^{(r)}(\bm{x})\,,\partial_{\bm{x}}\mathscr{g}_{s\,t}^{(r)} (\bm{y}|\bm{x})\,\right\rangle
		-\frac{1}{2}\operatorname{Tr}\left(\partial_{\bm{x}}\otimes\partial_{\bm{x}} \operatorname{A}_{t}(\bm{x})\mathscr{g}_{s\,t}^{(r)} (\bm{y}|\bm{x})\right) 
		= 0,
	\end{align}
\end{subequations}
where it now evolves with respect to $ s $ \emph{backwards in time}, i.e. for  values of $ s $ decreasing from $ t $ and
\begin{align}
	\bm{b}_{t}^{(r)}(\bm{x})=\bm{b}_{t}(\bm{x})-\frac{1}{p_{t}(\bm{x})}\partial_{\bm{x}}\cdot\big{(}\operatorname{A}_{t}(\bm{x})p_{t}(\bm{x})\big{)}.
	\nonumber
\end{align}
Alternatively, we may resort to the time reversal transformation
\begin{gather}
	t=\tf+\ti-t^{\prime}\hspace{1.0cm}\&\hspace{1.0cm}s=\tf+\ti-s^{\prime},
	\label{tr:tc}
	\shortintertext{so that}
	s\,\leq\,t \hspace{1.0cm}\Leftrightarrow \hspace{1.0cm} s^{\prime}\,\geq\,t^{\prime},
	\nonumber
\end{gather}
in order to associate to $ \mathscr{g}_{s\,t}^{(r)} $ a \emph{forward process} with transition probability density $  \tilde{\mathscr{g}}_{s^{\prime}\,t^{\prime}}$ specified by the identity
\begin{align}
	\tilde{\mathscr{g}}_{s^{\prime}\,t^{\prime}}(\bm{y}\big{|}\bm{x})=\mathscr{g}_{s\,t}^{(r)}(\bm{y}|\bm{x})
	\nonumber
\end{align}
holding for all $ \bm{x} $, $ \bm{y} $. It is then readily verified that inserting $ \tilde{\mathscr{g}}_{s^{\prime}\,t^{\prime}} $ in (\ref{tr:pair}) maps the "backward" Kolmogorov pair into the standard pair consisting of a forward Fokker--Planck equation and its adjoint.

\subsection{Consequences for Schr\"odinger's mass transport - general case }

It is instructive to rewrite  Schr\"odinger's mass transport equations (\ref{me:sy}) in terms of $ \mathscr{k}_{s\,t}^{(r)} $.  Some straightforward substitutions yield
\begin{subequations}
	\label{tr:sy}
	\begin{align}
		&\label{tr:sy1}
			w_{\mathfrak{1}}(\bm{x}_{\mathfrak{1}})=\varphi_{\mathfrak{1}}^{(r)}(\bm{x}_{\mathfrak{1}})
		\int_{\mathbb{R}^{d}}\mathrm{d}^{d}\bm{x}_{\mathfrak{0}}\,\varphi_{\mathfrak{0}}^{(r)}(\bm{x}_{\mathfrak{0}})
		\mathscr{g}_{\ti\,\tf}^{(r)}(\bm{x}_{\mathfrak{0}}|\bm{x}_{\mathfrak{1}}),
		\\
		&\label{tr:sy2}
			w_{\mathfrak{0}}(\bm{x}_{\mathfrak{0}})=\varphi_{\mathfrak{0}}^{(r)}(\bm{x}_{\mathfrak{0}})
		\int_{\mathbb{R}^{d}}\mathrm{d}^{d}\bm{x}_{\mathfrak{1}}\,
		\mathscr{g}_{\ti\,\tf}^{(r)}(\bm{x}_{\mathfrak{0}}|\bm{x}_{\mathfrak{1}})\varphi_{\mathfrak{1}}^{(r)}(\bm{x}_{\mathfrak{1}}),
	\end{align}
\end{subequations}
where we introduced 
\begin{align}
&	\varphi_{\mathfrak{0}}^{(r)}(\bm{x})=\frac{\varphi_{\mathfrak{0}}(\bm{x})}{p_{\ti}(\bm{x})},
\nonumber\\
&	\varphi_{\mathfrak{1}}^{(r)}(\bm{x})=\varphi_{\mathfrak{1}}(\bm{x})\,p_{\tf}(\bm{x}).
	\nonumber
\end{align}
Correspondingly, the equations for the $ \mathscr{g} $-harmonic function and its adjoint, eq.~(\ref{it:Born}), become
\begin{subequations}
	\label{tr:Born}
	\begin{align}
		&\label{tr:Bornf}
		h_{t}^{(r)}(\bm{x})\,\equiv\,\frac{\bar{h}_{t}(\bm{x})}{p_{t}(\bm{x})}=\int_{\mathbb{R}^{d}}\mathrm{d}^{d}\bm{x}_{\mathfrak{0}}\,\varphi_{\mathfrak{0}}^{(r)}(\bm{x}_{\mathfrak{0}})\mathscr{g}_{t_{\mathfrak{0}}\,t}^{(r)}(\bm{x}_{\mathfrak{0}}|\bm{x}),
		\\
		&\label{tr:Bornb}
		\bar{h}_{t}^{(r)}(\bm{x})\,\equiv\,h_{t}(\bm{x})\,p_{t}(\bm{x})=\int_{\mathbb{R}^{d}}\mathrm{d}^{d}\bm{x}_{\mathfrak{1}}\,
		\mathscr{g}_{t\,t_{\mathfrak{1}}}^{(r)}(\bm{x}|\bm{x}_{\mathfrak{1}})\,\varphi_{\mathfrak{1}}^{(r)}(\bm{x}_{\mathfrak{1}}).
	\end{align}
\end{subequations}
We see that the harmonic function and its adjoint exchange roles if we rephrase Schr\"odinger's mass transport equations in terms of the reversed transition probability density $\smash{\mathscr{g}_{t\,t_{\mathfrak{1}}}^{(r)} }$. 
The time reversal operation (\ref{tr:tc}) brings about a perhaps more interesting interpretation: we obtain forward dynamics 
with respect to the transition probability $ \tilde{\mathscr{g}}_{s^{\prime}\,t^{\prime}} $ whereas the boundary conditions $ w_{\mathfrak{0}} $, $ w_{\mathfrak{1}} $
exchange their roles.

\subsection{Consequences for Schr\"odinger's mass transport - detailed balance}

Following \S~4 of \citeTh[C]{Kol1936}, we now make two further assumptions:
\begin{itemize}
	\item 
		the transition probability density of the forward process is invariant under time translations
		\begin{align}
			\mathscr{g}_{t\,s}(\bm{x}|\bm{y})=	\mathscr{g}_{t-s}(\bm{x}|\bm{y})
			\nonumber
		\end{align}
		for all $ \bm{x} $,$ \bm{y} $;
	\item 
		the transition probability density of the forward process admits an invariant density $ p_{\star} $.
\end{itemize}
The hypotheses imply that the reversed transition probability specified by the invariant measure is then invariant under time translations. This property is inherited by the transition probability encoding the equivalent forward description
\begin{align}
	\tilde{\mathscr{g}}_{s^{\prime}\,t^{\prime}}(\bm{y}\big{|}\bm{x})=\tilde{\mathscr{g}}_{s^{\prime}-t^{\prime}}(\bm{y}\big{|}\bm{x})=\tilde{\mathscr{g}}_{t-s}(\bm{y}\big{|}\bm{x}).
	\nonumber
\end{align}
As a consequence (\ref{tr:K}) becomes
\begin{align}
	\label{tr:dbaux}
	p_{\star}(\bm{x})
	\,\tilde{\mathscr{g}}_{t-s}(\bm{y}|\bm{x})
	=
	\mathscr{g}_{t-s}(\bm{x}|\bm{y})
	p_{\star}(\bm{y}).
\end{align}
A stronger assumption is  the \emph{detailed balance condition}
\begin{gather}
	\label{tr:db}
	\tilde{\mathscr{g}}_{t-s}(\bm{y}|\bm{x})=\mathscr{g}_{t-s}(\bm{y}|\bm{x})
	\shortintertext{or, equivalently,}
	\nonumber
	p_{\star}(\bm{x})
	\,\mathscr{g}_{t-s}(\bm{y}|\bm{x})
	=
	\mathscr{g}_{t-s}(\bm{x}|\bm{y})\,p_{\star}(\bm{y}).
\end{gather}

The presence of detailed balance condition translates into an invariance property under time reversal of the solution of  Schr\"odinger's mass transport equations. More explicitly (\ref{tr:sy}) becomes
	\begin{align}
&		\nonumber
w_{\mathfrak{1}}(\bm{x})=\varphi_{\mathfrak{1}}(\bm{x}_{\mathfrak{1}})\,p_{\star}(\bm{x})
\int_{\mathbb{R}^{d}}\mathrm{d}^{d}\bm{y}\,\frac{\varphi_{\mathfrak{0}}(\bm{y})}{p_{\star}(\bm{y})}\,
\mathscr{g}_{\tf-\ti}(\bm{y}|\bm{x}),
\\
&
w_{\mathfrak{0}}(\bm{x})=\frac{\varphi_{\mathfrak{0}}(\bm{x})}{p_{\star}(\bm{x})}
\int_{\mathbb{R}^{d}}\mathrm{d}^{d}\bm{y}\,
\mathscr{g}_{\tf-\ti}(\bm{x}|\bm{y})\,\varphi_{\mathfrak{1}}(\bm{y})\,p_{\star}(\bm{y}).
		\nonumber
	\end{align}
Upon contrasting the above pair of equations with the time autonomous version of (\ref{me:sy}):
	\begin{align}
&\nonumber
	w_{\mathfrak{1}}(\bm{x})=\varphi_{\mathfrak{1}}(\bm{x})
	\int_{\mathbb{R}^{d}}\mathrm{d}^{d}\bm{y}\,
	\mathscr{g}_{\tf-\ti}(\bm{x}|\bm{y})\,
	\varphi_{\mathfrak{0}}(\bm{y}),
	\\
& \nonumber
	w_{\mathfrak{0}}(\bm{x})=\varphi_{\mathfrak{0}}(\bm{x})
	\int_{\mathbb{R}^{d}}\mathrm{d}^{d}\bm{y}\,\varphi_{\mathfrak{1}}(\bm{y})
	\mathscr{g}_{\tf-\ti}(\bm{y}|\bm{x}),
\end{align}
we arrive at the conclusion that  under the detailed balance hypothesis
(\ref{tr:db}) exchanging the boundary conditions $w_{\mathfrak{0}}\longleftrightarrow w_{\mathfrak{1}}$  maps a solution of the Schrödinger mass transport equation (\ref{me:sy}) into a solution according to the transformation law
\begin{align}
	(\varphi_{\mathfrak{0}},\varphi_{\mathfrak{1}})\rightarrow\left (\varphi_{\mathfrak{1}}\,p_{\star},\frac{\varphi_{\mathfrak{0}}}{p_{\star}}\right ).
	\nonumber
\end{align}
Under the same detailed balance hypothesis we verify that (\ref{tr:Born}) becomes
\begin{subequations}
	\label{tr:Borndb}
	\begin{align}
		&\label{tr:Borndb1}
		\frac{\bar{h}_{t}(\bm{x})}{p_{\star}(\bm{x})}=\int_{\mathbb{R}^{d}}\mathrm{d}^{d}\bm{y}\,\frac{\varphi_{\mathfrak{0}}(\bm{y})}{p_{\star}(\bm{y})}\mathscr{g}_{t-t_{\mathfrak{0}}}(\bm{y}|\bm{x})	,
		\\
		&\label{tr:Borndb2}
		h_{t}(\bm{x})\,p_{\star}(\bm{x})=\int_{\mathbb{R}^{d}}\mathrm{d}^{d}\bm{y}\,
		\mathscr{g}_{t_{\mathfrak{1}}-t}(\bm{x}|\bm{y})\,\varphi_{\mathfrak{1}}(\bm{y})\,p_{\star}(\bm{y}).
	\end{align}
\end{subequations}
If we now compare (\ref{tr:Borndb}) against the time autonomous limit of (\ref{it:Born}):
\begin{subequations}
	\label{tr:Bornta}
	\begin{align}
		&\label{tr:Bornta1}
		\bar{h}_{t}(\bm{x})=\int_{\mathbb{R}^{d}}\mathrm{d}^{d}\bm{y}\,\mathscr{g}_{t-t_{\mathfrak{0}}}(\bm{x}|\bm{y})	\varphi_{\mathfrak{0}}(\bm{y}),
		\\
		&\label{tr:Bornta2}
		h_{t}(\bm{x})=\int_{\mathbb{R}^{d}}\mathrm{d}^{d}\bm{y}\,\varphi_{\mathfrak{1}}(\bm{y})
		\mathscr{g}_{t_{\mathfrak{1}}-t}(\bm{y}|\bm{x}),
	\end{align}
\end{subequations}
we see that the time reversal operation (\ref{tr:tc}) respectively relates (\ref{tr:Borndb1}) to (\ref{tr:Bornta2}) and (\ref{tr:Borndb2}) to (\ref{tr:Bornta1}).
The interpretation is that  exchanging the boundary conditions $w_{\mathfrak{0}}\longleftrightarrow w_{\mathfrak{1}}$  occasions the transformation
	\begin{align}
		(h_{t},\overline{h}_{t})\longrightarrow\left (\bar{h}_{\ti+\tf-t}\,p_{\star},\frac{h_{\ti+\tf-t}}{p_{\star}}\right )
		\nonumber
	\end{align} 
so that the interpolating density (\ref{it:wt}) also transforms as
\begin{align}
	w_{t} \longrightarrow w_{\ti+\tf-t}
	\nonumber
\end{align}
for any $ t\,\in\,[\ti,\tf] $.  This is an important consequence of the property that Schr\"odinger calls \emph{reversibility}. In Schr\"odinger's words 
``\emph{thus the solution does not indicate any time direction. 
	If one exchanges $w_{0}$ with $w_{1}$, one obtains precisely the reverse evolution of $w_{t}(\bm{x})$
}''. This idea is further highlighted if  we couch the forward Kolmogorov equation associated to the solution of the optimal control problem in the form of the mass transport equation
\begin{align}
	\partial_{t}w_{t}(\bm{x})+\partial_{\bm{x}}\cdot \big{(}w_{t}(\bm{x})\bm{v}_{t}(\bm{x})\big{)}=0
	\nonumber
\end{align}
driven by the \emph{current velocity} \citeTh[C]{Nel1985}
\begin{align}
	\bm{v}_{t}(\bm{x})\,\equiv\,\lim_{\varepsilon\downarrow 0} \frac{\operatorname{E}\left (\bm{\xi}_{t+\varepsilon}-\bm{\xi}_{t-\varepsilon}\,\big{|}\bm{\xi}_{t}=\bm{x}\right )}{\varepsilon}=
	\bm{b}_{t}(\bm{x})+\operatorname{A}_{t}(\bm{x})\partial_{\bm{x}}h_{t}(\bm{x})-\frac{1}{2\,w_{t}(\bm{x})}\partial_{\bm{x}}\bigg{(}\operatorname{A}_{t}(\bm{x})w_{t}(\bm{x})\bigg{)}
	\label{tr:cv}
\end{align}
Under time reversal the current velocity transforms as 
\begin{align}
	\bm{v}_{t}(\bm{x}) \longrightarrow-\bm{v}_{\ti+\tf-t}(\bm{x})
	\nonumber
\end{align}
This means that the probability density is transported, in Schr\"odinger's words,  by a  ``\emph{diffusion current, which almost always almost exactly flowed in the direction of the concentration gradient (upward and not downward slope)}''. 
It is thus tempting  to interpret the discussion of \S~6 in Schr\"odinger's paper as a precursor of the optimal fluctuation theory later devised by Lars Onsager and Stefan Machlup  \citeTh[C]{OnMa1953}.

\section{Schr\"odinger's mass transport and Landauer's bound}
\label{sec:Landauer}

Thermodynamic processes at the micro-scale and below occur in highly fluctuating environments \citeTh[C]{Sek2010,Sei2012,PePi2020}. The recognition of this fact has driven the effort to extend thermodynamics to encompass closed and open systems evolving far from thermal equilibrium. The investigation of \emph{fluctuation relations} originating with the numerical observations of Denis Evans, Ezechiel Godert David Cohen and Gary Morriss \citeTh[C]{EvCoMo1993} and their theoretical explanation by Giovanni Gallavotti and Ezechiel Godert David Cohen \citeTh[C]{GaCo1995} played a pivotal role in this direction. Fluctuation relations are robust identities governing the statistics of 
thermodynamic quantifiers of the state of physical systems in and out thermal equilibrium. A major implication of the existence of fluctuation relations is the  extension of the Second Law of thermodynamics in order to properly take into account positive as well as negative fluctuations of the entropy production \citeTh[C]{Jar1997,Cro1997}.
This extension can be completely achieved when open systems are modeled by means of finite-dimensional Langevin dynamics \citeTh[C]{Kur1998,LeSp1999,MaReMo2000,JiQiQi2004}. In this latter context, a unified derivation of known fluctuation relations  follows from comparing in a mathematically rigorous manner how different choices of time-reversal transformations affect given forward  stochastic dynamics \citeTh[C]{ChGa2008}. 

\subsection{Elementary Langevin stochastic thermodynamics}

For instance, suppose that the physical system is described by a Markov process $ \left\{ \bm{\chi}_{t} \right\}_{t\,\geq\,0} $ taking values in a Euclidean space of dimension $2\,d$ and solution of
\begin{align}
	\mathrm{d}\bm{\chi}_{t}=\left(\mathsf{J}
	-\mathsf{S}\right)\partial_{\bm{\chi}_{t}}H_{t}(\bm{\chi}_{t})
	\,\mathrm{d}t+\sqrt{ \frac{2}{\beta}}\mathsf{S}^{1/2} \mathrm{d}\bm{w}_{t}.
	\label{Landauer:sde}
\end{align}  
Here $H_{t}$ is a scalar, possibly time dependent, function called the Hamiltonian, and  $ \mathsf{J} $ and $ \mathsf{S} $ are an antisymmetric and a symmetric positive definite matrix, respectively. 
The drift in (\ref{Landauer:sde}) is the sum of $ \mathsf{J}\partial_{\bm{x}}H_{t} $, an incompressible  component, and a gradient $ -\mathsf{S}\partial_{\bm{x}}H_{t} $. If the Hamiltonian is time independent $ H_{t}\,\equiv\,H $, the incompressible component preserves $ H $ and for this reason it is referred to as the ``conservative'' component also in the general case. Similarly, the gradient component is referred to as the ``dissipative'' component as in the time autonomous case it drives the solution towards the minimum of $ H $ (if it exists).
The Wiener differential $\mathrm{d}\bm{w}_{t}$ in (\ref{Landauer:sde}) models thermal exchanges between the system and an infinite environment at temperature~$\beta^{-1}$.

The kinematics of (\ref{Landauer:sde}) are chosen such to satisfy the \emph{Einstein relation} \citeTh[C]{ChGa2008}. This means that under additional hypotheses on the Hamiltonian (e.g. time independent, confining)  the probability measure generated by (\ref{Landauer:sde}) converges for large times towards a unique Boltzmann equilibrium:
\begin{equation*}
	p_{\infty}(\bm{x})=\frac{e^{-\beta\,H(\bm{x})}}{Z},
	\qquad
 	Z= \int_{\mathbb{R}^{2\,d}}\mathrm{d}\bm{x}\,e^{-\beta\,H(\bm{x})}.
\end{equation*}
More generally  for any finite time interval $ [\ti\,,\tf] $ we may interpret the Stratonovich stochastic differential (see e.g. \citeTh[C]{Sch2010})
\begin{align}
	H_{\tf}(\bm{\chi}_{\tf})-H_{\ti}(\bm{\chi}_{\ti})=\int_{\ti}^{\tf}\mathrm{d}t\,(\partial_{t}H_{t})(\bm{\chi}_{t})+\int_{\ti}^{\tf}\left \langle\,\mathrm{d}\bm{\chi}_{t}\,,\partial_{\bm{\chi}_{t}}H_{t}(\bm{\chi}_{t})\,\right\rangle
	\label{Landauer:1st}
\end{align}
as a stochastic embodiment of the first law of thermodynamics \citeTh[C]{Sek1998}. In particular, we identify 
\begin{align}
	Q_{\tf,\ti}=-\int_{\ti}^{\tf}\left \langle\,\mathrm{d}\bm{\chi}_{t}\,,\partial_{\bm{\chi}_{t}}H_{t}(\bm{\chi}_{t})\,\right\rangle
	\nonumber
\end{align}
as the heat released by an individual realization of the system evolution for $ t\,\in \,[\ti\,,\tf] $. A straightforward application of stochastic calculus then shows that
the expectation value of the heat can always be couched into the form (see e.g. \citeTh[C]{AuGaMeMoMG2012,Gaw2013})
\begin{align}
	\operatorname{E}Q_{\tf,\ti}=\int_{\mathbb{R}^{2\,d}}\mathrm{d}\bm{x}\,\frac{p_{\tf}(\bm{x})\ln p_{\tf}(\bm{x})-p_{\ti}(\bm{x})\ln p_{\ti}(\bm{x})}{\beta}
	+\int_{\ti}^{\tf}\mathrm{d}t\int_{\mathbb{R}^{2\,d}}\mathrm{d}\bm{x}\,p_{t}(\bm{x})\left\|\partial_{\bm{x}}\left (H_{t}(\bm{x})+\frac{1}{\beta}\ln p_{t}(\bm{x})\right )\right\|_{\mathsf{S}}^{2}
	\nonumber
\end{align}
where $ p_{t} $ is the probability density of states of the system at time $ t $.
Of the two terms on the right-hand side, the first one is a non-sign-definite time-boundary term. It coincides with minus the variation of the Gibbs-Shannon entropy 
\begin{align}
	S_{t}=-\int_{\mathbb{R}^{2\,d}}\mathrm{d}\bm{x}\,p_{t}(\bm{x})\ln p_{t}(\bm{x}).
	\nonumber
\end{align}
From the thermodynamic point of view, we  interpret it as measuring the system entropy variation in consequence of the transition.
The second term on the right-hand side is positive definite and vanishes identically only at equilibrium. Already these elementary phenomenological considerations suggest  the interpretation of the second term as the average entropy production during the thermodynamic transition. The analysis of the fluctuation relation between the probability measure of the process (\ref{Landauer:sde}) and that of the process constructed by treating the dissipative and conservative components of the drift as respectively even and odd under the physically natural choice of time-reversal transformation ultimately validates the identification. The result of the analysis \citeTh[C]{MaReMo2000,ChGa2008} is the identity
\begin{align}
\int_{\ti}^{\tf}\mathrm{d}t\int_{\mathbb{R}^{2\,d}}\mathrm{d}\bm{x}\,p_{t}(\bm{x})\left\|\partial_{\bm{x}}\left (H_{t}(\bm{x})+\frac{1}{\beta}\ln p_{t}(\bm{x})\right )\right\|_{\mathsf{S}}^{2}=\frac{1}{\beta}\int \mathrm{d}\mathrm{P} \ln\frac{\mathrm{d}\mathrm{P}}{\mathrm{d}\mathrm{P}^{(r)}\circ R}\,\equiv\,\frac{1}{\beta}D_{\scriptscriptstyle{KL}}(\mathrm{P}\|\mathrm{P}^{(r)}\circ R)
	\label{Landauer:KL}
\end{align}
proving that the positive definite component in the average heat coincides with Kullback--Leibler between the measure of  forward process $ \mathrm{P} $ generated by (\ref{Landauer:sde}) and the image-measure $ \mathrm{P}^{(r)}\circ R $  by path reversal $ R $ \citeTh[C]{ChGa2008,Che2018} of the time-reversed process $  \mathrm{P}^{(r)}$. Combining (\ref{Landauer:KL}) with the observation that the change of entropy of environment is related to the average heat release, we arrive at the expression of the average value of the Second Law in the framework  of Langevin thermodynamics:
\begin{align}
	\Delta S_{Tot}=\beta\operatorname{E}Q_{\tf,\ti}+S_{\tf}-S_{\ti}=D_{\scriptscriptstyle{KL}}(\mathrm{P}\|\mathrm{P}^{(r)}\circ R )
	\,\geq\, 0.
	\label{Landauer:2nd}
\end{align}
The total entropy production during an arbitrary transition can only increase. Conversely, only transitions governed by a probability measure invariant under time-reversal do not occasion an increase of the total entropy.

\subsection{Landauer's principle in the context of Langevin dynamics}

Once expressed in the form (\ref{Landauer:2nd}), the Second Law of stochastic thermodynamics is closely related to the Landauer's principle \citeTh[C]{Lan1961}. The principle  states that the erasure of one bit of information performed in a thermal environment produces on average a heat release no smaller than  $  \beta^{-1}\,\ln 2 $. The bound is obtained in the quasi-static limit. The existence  of a strictly positive lower bound to the heat release during erasure  indicates a fundamental minimum cost that must be paid to run any computing process. It is, however, worth emphasizing here that thermodynamic processes in fluctuating thermal environments are described by stochastic quantities. Hence a  probabilistic formulation of Landauer’s principle other than existence on average may play an important role to determine the actual fundamental cost of computing \citeTh[C]{DiLu2009}. 
Nevertheless, a natural question to ask concerns corrections to the $  \beta^{-1}\,\ln 2 $ heat release estimate occasioned by a \emph{finite-time} transition \citeTh[C]{AuGaMeMoMG2012}. In the context of Langevin thermodynamics, the corresponding mathematical problem is pictorially described in Fig.~\ref{fig:Landauer}.
A stored bit of information is modeled by a probability density at time $ \ti $ having a single sharp maximum for instance  to the right of the origin. 
Finite-time erasure consists of steering the probability density so that at time $ \tf $ it acquires two symmetric maxima around the origin. The physically relevant cost to minimize
\emph{with respect to the drift} is the Kullback--Leibler divergence on the right-hand side of (\ref{Landauer:2nd}). Conceptually, the resulting optimal control problem is very close to the one leading to Schr\"odinger's mass transport.
The important difference resides in the definition of the cost. In Schr\"odinger's case the cost of a transition depends upon the, in principle arbitrary, choice of the reference process.  This arbitrariness is not present in the stochastic thermodynamics formulation of Landauer's principle. The  Kullback--Leibler divergence appearing in the Second Law of thermodynamics (\ref{Landauer:KL}) is fully specified in terms of the drift and diffusion of the control process. Furthermore, it has the interpretation of total entropy production during a thermodynamic transition and it only vanishes at equilibrium. This fact becomes especially evident in the absence of a conservative component  in the drift of (\ref{Landauer:sde}) ($ \mathsf{J}=0 $) and if $ \mathsf{S} $ is strictly positive definite. In such a case, the Kullback--Leibler divergence (\ref{Landauer:KL}) admits the simple expression
\begin{align}
	D_{\scriptscriptstyle{KL}}(\mathrm{P}\|\mathrm{P}^{(r)}\circ R )=\int_{\ti}^{\tf}\mathrm{d}t\,\int_{\mathbb{R}^{d}}\mathrm{d}^{d}x\,\left\|\bm{v}_{t}(\bm{x})\right\|_{\operatorname{S}^{-1}}^{2} w_{t}(\bm{x}),
	\nonumber
\end{align}
where as before $ w_{t} $ is the probability density of the forward process and the vector field  $  \bm{v}_{t}$ is the current velocity (\ref{tr:cv})
\begin{align}
	\bm{v}_{t}(\bm{x})=\mathsf{S}\partial_{\bm{x}}\left(H_{t}(\bm{x})+\frac{1}{\beta}\ln p_{t}(\bm{x})\right).
	\nonumber
\end{align}
At equilibrium the current velocity vanishes and no steering between distinct states is possible. We can instead naturally rephrase the optimization problem using the current velocity as control to minimize the erasure cost. 
As a consequence \citeTh[C]{AuGaMeMoMG2012}, the optimal control equations for the mean average dissipation in a thermodynamic transition coincide with those of a classical optimal mass transport \citeTh[C]{Vil2009}. Furthermore, Schr\"odinger's mass transport problem with reference process a free diffusion ($ \bm{b}_{t}=0 $ in (\ref{vp:HJB})) coincides with  a viscous regularization of the classical mass transport \citeTh[C]{PMG2013}. Hence the conceptual proximity between the two optimal control process becomes in this special case a quantitative relation. More generally, the same quantitative relation holds for micro-scale processes described by the Langevin--Smoluchowski, overdamped, approximation of (\ref{Landauer:sde}) \citeTh[C]{AuGaMeMoMG2012}. 

The general interest of a quantitative relation between the optimal control problems associated to Schr\"odinger's mass transport and Landauer's principle is motivated by the following considerations. 
Ongoing experiments, e.g. \mbox{\citeTh[C]{DaPeBaCiBe2021}}, are pushing towards a better understanding of the cost of information processing by nano-scale machines, natural or artificial. 
At the nano-scale  inertial interactions cannot be neglected. 
In such a case, even highly stylized models of the  dynamics neglecting quantum effects require the use of the full-fledged Langevin-Kramers dynamics \citeTh{Kra1940, Zwa2001}. 
A~direct quantitative correspondence with Schr\"odinger's stochastic optimal control problem  is no longer immediately evident. 
Nevertheless, the Langevin-Smoluchowski limit remains a stepping stone for analytical investigations based on multiscale perturbation theory \citeTh[C]{PMGSc2014}. 
In addition, the solution of Landauer's optimal control problem in the Langevin-Smoluchowski limit is also the basis for the analysis of erasure when the bit of information is conceptualized as a macro-state specified by coarse graining the measure of a  physical system with microscopic dynamics governed by (\ref{Landauer:sde}) \citeTh[C]{PrEhBe2020}.
Thus, in a broad sense, Schr\"odinger's vision of an optimal stochastic mass transport problem between target states offers a powerful theoretical framework for conceptualizing transitions in stochastic thermodynamics.

\section{Conclusion}

Schrödinger's 1931 paper ``On the Reversal of the Laws of Nature''  appeared amid the debate on the interpretation of quantum mechanics. 
The paper triggered  manifold developments of the theory of stochastic processes, directly or indirectly related to the still ongoing effort  to shed light on the connections and the physical origins 
of the differences between classical statistical physics and probability theory on one side and quantum mechanics on the other.  
The~advent of micro- and nano-scale technology in the last decades has made urgent the need for a deeper understanding of thermodynamic processes which occur in finite time and  involve physical quantities described by inherently fluctuating quantities. 
Forging a theoretical framework adapted to  match this challenge is a task that a significant part of the community working in theoretical and mathematical physics has endeavored to undertake. 
In~our attempt to contribute to this collective effort, we found inspiration and guidance in reading, possibly from a slightly novel perspective,  Schr\"odinger's ``On the Reversal of the Laws of Nature''. 
We therefore decided to offer our translation and commentary of  ``On the Reversal of the Laws of Nature'' with the hope that, as it was for us, other colleagues may find in it a source of ideas to open new paths both in  their research and teaching activity.

\section{Acknowledgments}

We warmly thank two anonymous reviewers for their careful reading of our manuscript and many very useful suggestions. Their  feedback allowed us to greatly improve the quality of the translation and of the paper overall.
The authors are also glad to acknowledge useful comments and encouragement from Angelo Vulpiani, John Bechhoefer, Hugo Touchette, Massimo Cencini, Brecht Donvil, and Ruben Pasmanter. Special thanks go to Michael McAuley who helped us to improve the language quality of the text.
R.C. is supported by the French National Research Agency through the projects QTraj (ANR-20-CE40-0024-01), RETENU (ANR-20-CE40-0005-01), and ESQuisses (ANR-20-CE47- 0014-01). 
\appendix

\renewcommand{\refname}{References for Commentary [C]}
\bibliographyTh{reversibility_comments} 
\bibliographystyleTh{abbrv}


\end{document}